\documentclass[structabstract]{aa}

\usepackage{graphicx}
\usepackage{subfigure}
\usepackage{natbib}
\usepackage{float}
\usepackage{hyperref}
\usepackage{amsmath}
\usepackage{supertabular}
\usepackage{longtable}
\usepackage{txfonts}

\newcommand{\tabincell}[2]{\begin{tabular}{@{}#1@{}}#2\end{tabular}}

\def\Jtwo  {J2052$+$3635}  
\def\Jfour {J2044$+$4005}  
\def\Jsix  {J2046$+$4106}  %
\def\Jnine {J2049$+$4118}  %
\def\hho     {H$_2$O}

\def\VLSR    {$V_{\rm LSR}$}

\def\kms     {km~s$^{-1}$}
\def\masy    {mas~yr$^{-1}$}
\def\mjybeam {mJy~beam$^{-1}$}
\def\jybeam  {Jy~beam$^{-1}$}

\def\msun    {$M_{\odot}$}
\def\Lsun    {$L_{\odot}$}
\def\Lbol    {$L_{\rm bol}$}
\def\dLsun   {$(d/{\rm kpc})^2~L_{\odot}$}

\def\Teff    {$T_{\rm eff}$}
\def\Rrp     {$R_{\rm rp}$}
\def\Rop     {$R_{\rm op}$}
\def\Trp     {$T_{\rm rp}$}
\def\Top     {$T_{\rm op}$}

\def\dx      {$\Delta x$}
\def\dy      {$\Delta y$}
\def\ra      {$\alpha_{\rm J2000}$}
\def\dec     {$\delta_{\rm J2000}$}

\def\h       {\ifmmode{^{\rm h}}\else{$^{\rm h}$}\fi}
\def\m       {\ifmmode{^{\rm m}}\else{$^{\rm m}$}\fi}
\def\s       {\ifmmode{^{\rm s}}\else{$^{\rm s}$}\fi}
\def\deg     {\ifmmode{^{\circ}}\else{$^{\circ}$}\fi}
\def\decdeg  {\ifmmode{{\rlap.}^{\circ}} \else ${\rlap.}^{\circ}$\fi}
\def\decs    {\ifmmode{{\rlap.}^{\rm s}} \else ${\rlap.}^{\rm s}$\fi}
\def\decas   {\ifmmode{{\rlap.}{''}}\else{${\rlap.}{''}$}\fi}

\def\Ts    {\ifmmode{\Theta_s}\else{$\Theta_s$}\fi}
\def\Tdot  {\ifmmode{d\Theta\over dR}\else{$d\Theta\over dR$}\fi}
\def\Rs    {\ifmmode{R_s}\else{$R_s$}\fi}
\def\To    {\ifmmode{\Theta_0}\else{$\Theta_0$}\fi}
\def\Ro    {\ifmmode{R_0}\else{$R_0$}\fi}

\def\Vo    {\ifmmode {V^{Std}_\odot}\else {$V^{Std}_\odot$}\fi}
\def\Uo    {\ifmmode {U^{Std}_\odot}\else {$U^{Std}_\odot$}\fi}
\def\Wo    {\ifmmode {W^{Std}_\odot}\else {$W^{Std}_\odot$}\fi}
\def\VH    {\ifmmode {V^H_\odot}\else {$V^H_\odot$}\fi}
\def\UH    {\ifmmode {U^H_\odot}\else {$U^H_\odot$}\fi}
\def\WH    {\ifmmode {W^H_\odot}\else {$W^H_\odot$}\fi}
\def\V     {\ifmmode {V_\odot}\else {$V_\odot$}\fi}
\def\U     {\ifmmode {U_\odot}\else {$U_\odot$}\fi}
\def\W     {\ifmmode {W_\odot}\else {$W_\odot$}\fi}
\def\VGC   {\ifmmode {V_\odot^{GC}}\else {$V_\odot^{GC}$}\fi}
\def\UGC   {\ifmmode {U_\odot^{GC}}\else {$U_\odot^{GC}$}\fi}
\def\WGC   {\ifmmode {W_\odot^{GC}}\else {$W_\odot^{GC}$}\fi}

\def\Vs    {\ifmmode {V_s}\else {$V_s$}\fi}
\def\Us    {\ifmmode {U_s}\else {$U_s$}\fi}
\def\Ws    {\ifmmode {W_s}\else {$W_s$}\fi}

\def\Vsbar {\ifmmode {\overline{V_s}}\else {$\overline{V_s}$}\fi}
\def\Usbar {\ifmmode {\overline{U_s}}\else {$\overline{U_s}$}\fi}
\def\Wsbar {\ifmmode {\overline{W_s}}\else {$\overline{W_s}$}\fi}

\def\mux    {\ifmmode {\mu_x}\else {$\mu_x$}\fi}
\def\muy    {\ifmmode {\mu_y}\else {$\mu_y$}\fi}
\def\mura   {\ifmmode {\mu_{\alpha}}\else {$\mu_{\alpha}$}\fi}
\def\mude   {\ifmmode {\mu_{\delta}}\else {$\mu_{\delta}$}\fi}

\newcommand{\HII}{\mbox{H\,\textsc{ii}}}%

\begin{document}

\title{The Distance and Size of the Red Hypergiant NML Cyg\ from VLBA
and VLA Astrometry}

\author{
B. Zhang\inst{1,2},
M. J. Reid\inst{3}, 
K. M. Menten\inst{1},
X. W. Zheng\inst{4},
A. Brunthaler\inst{1,5}}

\institute{
Max-Plank-Institut f\"ur Radioastronomie,
Auf dem H\"ugel 69, 53121 Bonn, Germany\\
\email{bzhang@mpifr.de}
\and
Shanghai Astronomical Observatory, Chinese Academy of Sciences,
80 Nandan Road, Shanghai 200030, China
\and
Harvard-Smithsonian Center for Astrophysics,
60 Garden Street, Cambridge, MA 02138, USA
\and
Department of Astronomy, Nanjing University,
22 Hankou Road, Nanjing 210093, China
\and
National Radio Astronomy Observatory, Socorro, NM 87801, USA}

\date{Received / Accepted}
\titlerunning{Distance and size of NML Cyg}
\authorrunning{B. Zhang et al.}

\abstract
   {The red hypergiant NML Cyg has been assumed to be at part of
   the Cyg OB2 association, although its distance has never been
   measured directly.  A reliable distance is crucial to study the
   properties of this prominent star.  For example, its luminosity, and
   hence its position on the H-R diagram, is critical information to
   determine its evolutionary status. In addition, a detection of the
   radio photosphere would be complementary to other methods of
   determining the stellar size.}
   {We aim to understand the characteristics of NML Cyg with direct
   measurements of its absolute position, distance, kinematics, and
   size.}
   {We observe circumstellar 22 GHz \hho\ and 43 GHz SiO masers with the
   Very Long Baseline Array to determine the parallax and proper motion
   of NML Cyg.  We observe continuum emission at 43 GHz from the radio
   photosphere of NML Cyg with the Very Large Array.}
   {
   We measure the annual parallax of NML Cyg to be 0.620 $\pm$ 0.047
   mas, corresponding to a distance of $1.61^{+0.13}_{-0.11}$ kpc. The
   measured proper motion of NML Cyg is \mux\ = $-1.55\pm 0.42$ \masy\
   eastward and \muy\ = $-4.59 \pm 0.41$ \masy\ northward.  Both the
   distance and proper motion are consistent with that of Cyg OB2,
   within their joint uncertainty, confirming their association.  Taking
   into consideration molecular absorption signatures seen toward NML
   Cyg, we suggest that NML Cyg lies on the far side of the Cyg OB2
   association.  The stellar luminosity revised with our distance brings
   NML Cyg significantly below the empirical luminosity limit for a red
   supergiant.  We partially resolve the radio photosphere of NML Cyg at
   43 GHz and find its diameter is about 44 mas, suggesting an optical
   stellar diameter of 22 mas, if the size of radio photosphere is 2
   times the optical photosphere.  Based on the position of
   circumstellar SiO masers relative to the radio photosphere, we
   estimate the absolute position of NML Cyg at epoch 2008.868 to be
   \ra\ = 20\h 46\m 25\decs5382 $\pm$ 0\decs0010, \dec\ = 40\deg
   06\arcmin59\decas379 $\pm$ 0\decas015.  The peculiar motions of NML
   Cyg, the average of stars in Cyg OB2, and four other star-forming
   regions rules out that an expanding ``Str\"omgren sphere'' centered
   on Cyg OB2 is responsible for the kinematics of the Cygnus X region.
   }
   {}


\maketitle

\section{Introduction}

NML Cyg is one of the most massive and luminous red hypergiants.  It is
a semi-regular star with period $\approx$ 1000 days
\citep{1988PZ.....22..882D, 1997ApJ...481..420M} and has been suggested
to be associated with  Cyg OB2, possibly the largest stellar association
in the Galaxy \citep{1982A&A...108..412H,1983ApJ...267..179M}.  Previous
distance estimates for NML Cyg have been mainly based on the assumption
that it is at the distance with Cyg OB2.   Even so, the distance to Cyg
OB2 is hard to determine accurately by photometry and spectroscopy, due
to difficulty in calibrating spectral types and luminosities.  Previous
investigators have derived distance moduli of 10.8 -- 11.6
\citep{1954ApJ...119..344J, 1966PROE....5..111R, 1973ApJ...180L..35W,
1981ApJ...250..645A, 1991MNRAS.249....1T, 1991AJ....101.1408M,
2003ApJ...597..957H, 2005A&A...438.1163K}, corresponding to distances of
1.45 -- 2.10 kpc to Cyg OB2.  These values are marginally
different from $1.22 \pm 0.3$ kpc to NML Cyg determined by
\citet{2001ApJ...555..405D}, based on the comparison of Doppler
velocities and proper motions of two discrete dust shells around NML Cyg
measured with interferometry at 11 $\mu$m over a period about 6 yr.

Adopting a distance of 1.74 $\pm$ 0.2 kpc \citep{1991AJ....101.1408M},
\citet{2009ApJ...699.1423S} estimated NML Cyg's minimum bolometric
luminosity to be $3.15 \pm 0.74 \times~10^5$ \Lsun, which is similar to
that of other red hypergiants and places it near the empirical
upper-luminosity boundary in the Hertzsprung-Russell (H-R) diagram
\citep{2006AJ....131..603S}.  However, the estimated luminosity of a
star depends on the square of its distance. Thus, an accurate distance
is crucial to derive a reliable luminosity and the location on the H-R
diagram.  Recently, trigonometric parallaxes of circumstellar maser
sources (\hho\ or SiO) surrounding red hypergiants with VLBI
phase-referencing have been measured with accuracies of 20 -- 80 $\mu$as
\citep{2008PASJ...60.1007C, 2012ApJ...744...23Z, 2010ApJ...721..267A}.
We carried out a program to measure the trigonometric parallax and
proper motion of masers in the circumstellar envelope (CSE) of NML Cyg
with the National Radio Astronomical Observatory's\footnote{The national
Radio Astronomy Observatory is a facility of the National Science
Foundation operated under cooperative agreement by Associated
Universities, Inc} (NRAO's) Very Long Baseline Array (VLBA).

In addition to distance, another fundamental stellar parameter is size.
NML Cyg's high mass-loss rate results in a dense circumstellar envelope,
and the star is faint and hard to observe at visual wavelengths due to
high extinction.  However, it is extremely luminous in the infrared.
\citet{1986ApJ...302..662R} used infrared speckle interferometric
observations of NML Cyg and a simple model of multiple shells to derive
an inner radius for its dust shell of 45 $\pm$ 12 mas.  They also
obtained the star's effective temperature, \Teff\ $\approx$ 3250 K,
which is appropriate for an M6 III star \citep{1967ApJ...149..345J}.
\citet{1997ApJ...481..420M} conducted infrared interferometry at 11.15
$\mu$m, yielding evidence for multiple dust shells and asymmetric dust
emission around NML Cyg.  They obtained better fits to all data by
assuming a \Teff\ $\approx$ 2500 K, using a model
\citep{1983MNRAS.202..767R}.  \citet{2001A&A...369..142B} presented
diffraction-limited 2.13 $\mu$m observations with 73 mas resolution and
obtained a bolometric flux of $F_{bol} = 3.63 \times 10^9$ W~m$^{-2}$,
corresponding to a stellar luminosity of \Lbol\ = $1.13 \times 10^5
\cdot $ \dLsun.  Using the spectral energy distribution (SED) from 2 to
50 $\mu$m, they derived a stellar diameter of 16.2 mas assuming \Teff\ =
2500 K. Clearly, a direct detection of the star would be complementary
to the methods mentioned above.  Radio continuum emission from the evolved
star's photosphere can be imaged with the Very Large Array (VLA), using
\hho\ or SiO masers as a phase reference \citep{1990ApJ...360L..51R,
1997ApJ...476..327R, 2007ApJ...671.2068R, 2012ApJ...744...23Z}. Such
observations allow us to directly measure the size of NML Cyg's radio
photosphere and, using the observed flux density and our measured
distance, its effective temperature.

\section{Observations and data reduction} \label{sec:obs}

\subsection{VLBA phase-referencing observations}

We conducted VLBI phase-referencing observations of NML Cyg\ at 22 and
43 GHz with the VLBA under program BZ036 with 7-hour tracks on 2008 May
25, August 13 and November 8, and 2009 January 27 and May 3.  This
sequence samples the peaks of the sinusoidal trigonometric parallax
curve and has low correlation coefficients among the parallax and
proper motion parameters.  We observed several extragalactic radio
sources as potential background references for parallax solutions.  We
alternated between two $\approx$ 16 min blocks at 22 and 43 GHz.  Within
a block, the observing sequence was NML Cyg, \Jtwo, NML Cyg, \Jfour, NML
Cyg, \Jsix, NML Cyg, \Jnine.  We switched between the maser target and
background sources every 40~s for 22 GHz and 30~s for 43 GHz, typically
achieving 30~s for 22 GHz and 20~s for 43 GHz of on-source data. 

We placed observations of two strong sources 3C345 (J1642+3948) and
3C454.3 (J2253+1608) near the beginning, middle, and end of the
observations in order to monitor delay and electronic phase differences
among the intermediate-frequency bands.  The rapid-switching
observations employed four adjacent bands of 8~MHz bandwidth and
recorded both right and left circularly polarized signals.  The \hho\
and SiO masers were contained in the second band centered at a Local
Standard of Rest (LSR) velocity of $-$7 \kms.  In order to do
atmospheric delay calibration, we placed ``geodetic" blocks before and
after our phase-reference observations.  These data were taken in left
circular polarization with eight 8~MHz bands that spanned 480~MHz of
bandwidth between 22.0 and 22.5~GHz; the bands were spaced in a
``minimum redundancy configuration" to uniformly sample, as best as
possible, all frequency differences.

\begin{figure}
  \centering
  \includegraphics[angle=0,scale=0.57]{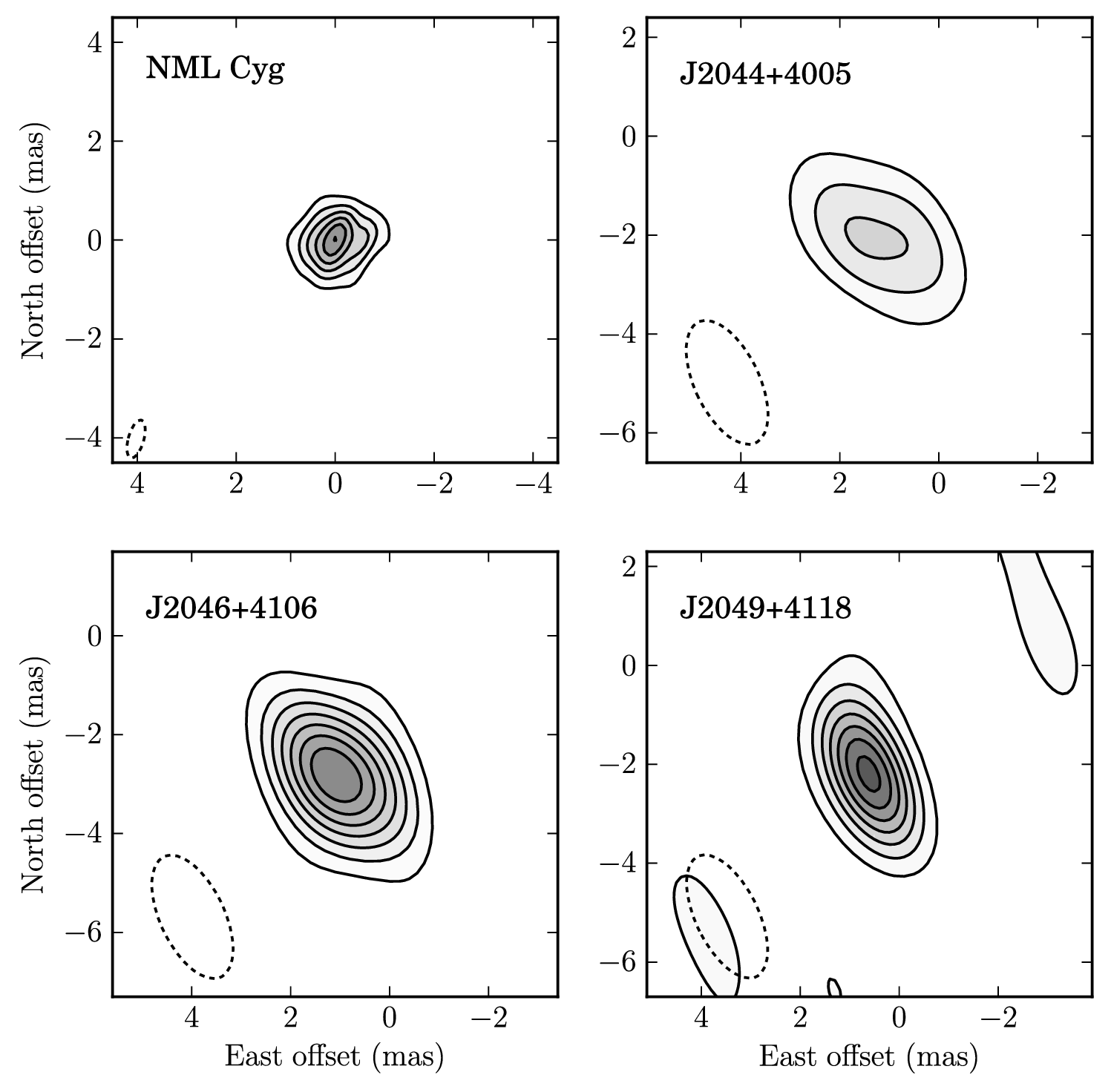}
  \caption{
Images of the \hho\ maser reference spot and the extragalactic radio
sources used for the parallax measurements of NML Cyg at the middle
epoch (2008 November 08).  Source names are in the upper left corner and
the restoring beam (dotted ellipse) is in the lower left corner of each
panel.  Contour levels are spaced linearly at 3 \jybeam\ for NML Cyg and
1.5 \mjybeam\ for the background sources.
  \label{fig:maser_qso}}
\end{figure}

The data were correlated in two passes with the VLBA correlator in
Socorro, NM.  One pass generated 16 spectral channels for all the data
and a second pass generated 256 spectral channels, but only for the
single (dual-polarized) frequency band containing the maser signals,
giving a velocity resolution of 0.42 \kms\ for 22 GHz and 0.21 \kms\ for
43 GHz, respectively.  Generally, the data calibration was performed
with the NRAO Astronomical Image Processing System (AIPS), following
similar procedures as described in \citet{2009ApJ...693..397R}.  We used
an \hho\ maser spot at a \VLSR\ of 5.22 \kms\ and a SiO maser spot at a
\VLSR\ of $-5.70$ \kms\ as phase-reference sources, because
they are considerable stronger than the background source and could be
detected on individual baselines in the available on-source time. We
imaged the calibrated data with the AIPS task IMAGR;
Fig.~\ref{fig:maser_qso} shows example images at 22 GHz from the
middle epoch.  Table~\ref{tab:src_pos} lists the positions, intensities,
source separations, LSR velocity of the reference maser spot and
synthesized beam sizes.

\begin{table*}
  \footnotesize
  \caption[]{Positions and Brightnesses \label{tab:src_pos}}
  \begin{center}
  \begin{tabular}{cccccccc}
\hline \hline
Source      &  R.A. (J2000) & Dec. (J2000)   & $S_p$      &  $\theta_{sep}$   & P.A. & \VLSR & Beam \\
            & (h~~~m~~~s)   & (\degr~~~\arcmin~~~\arcsec) & (Jy/beam) &  (\degr) & (\degr)  & (\kms) & (mas~~mas~~\degr) \\
 (1)        &   (2)         &   (3)          & (4)        & (5)       & (6)      & (7)      & (8)    \\
\hline
22 GHz&&& &&& & \\
NML Cyg.......& 20 46 25.5444  & 40 06 59.383  &  16 $-$ 77     &  ...  & ...     & 5.22 & 0.6 $\times$ 0.3 @ $-$08\\
\Jtwo....     & 20 52 52.0550  & 36 35 35.300  &   0.050        &  3.7  & $-$19   & ...  & 2.7 $\times$ 1.4 @ $+$31\\
\Jfour....    & 20 44 11.0877  & 40 05 36.360  &   0.005        &  0.4  & $+$87   & ...  & 2.7 $\times$ 1.3 @ $+$25\\
\Jsix....     & 20 46 21.8414  & 41 06 01.107  &   0.014        &  1.0  & $-$01   & ...  & 2.7 $\times$ 1.2 @ $+$24\\
\Jnine....    & 20 49 43.9548  & 41 18 14.583  &   0.011        &  1.3  & $+$28   & ...  & 2.7 $\times$ 1.2 @ $+$26\\
 &&& &&& & \\
43 GHz&&& &&& & \\
NML Cyg.......& 20 46 25.5378  & 40 06 59.413  &     3.4        &  ...  & ...     &$-5.70$& 0.3 $\times$ 0.1 @ $-$07\\
\Jsix....     & 20 46 21.8414  & 41 06 01.107  &   0.017        &  1.0  & $-$01   & ...  & 0.7 $\times$ 0.4 @ $-$25\\
\hline
  \end{tabular}
  \end{center}
\tablefoot{
The fourth and seventh columns give the peak brightnesses
($S_p$) and \VLSR\ of reference maser spot.  The fifth and sixth columns
give the separations ($\theta_{sep})$ and position angles (P.A.) east of
north of the background sources relative to the maser.  The last column gives the
FWHM size and P.A. of the Gaussian restoring beam.  Calibrator \Jtwo\ is
from the VLBA Calibrator Survey (VCS) by \citet{2007AJ....133.1236K}.
The other calibrators are from the VLA program BZ035.
}
\end{table*}

\subsection{VLA observation for radio photosphere}

In addition to the VLBA observations, we carried out an observation of
the radio photosphere and maser emission toward NML Cyg\ at 43 GHz with
the VLA in its largest (A) configuration under program AZ178 on 2008
December 20.  We used  a dual intermediate frequency band setup in
continuum mode with
a narrow (3.125 MHz) band centered at \VLSR\ of $-18$ \kms\ for the $v =
1, J = 1 \to 0$ SiO maser line (rest frequency of 43122.027 MHz) and a
broad (50 MHz) band centered $\approx50$ MHz above in a
line-free portion of the spectrum. Additionaly, we observed the SiO maser emission
in two sub-bands centered at \VLSR\ of $-18$ and $-5$ \kms\ at high spectral
resolution (channel spacing of 97.6563 kHz and bandwidth of 6.25 MHz)
with several scans in spectral-line mode interspersed among the
dual-band continuum observations. We also observed \hho\ maser emission
(rest frequency of 22235.080 MHz) centered at \VLSR\ of 5 \kms\ with a
channel spacing of 48.8281 kHz and bandwidth of 6.25 MHz in a snapshot
mode.

A typical observing unit for our continuum observations consisted of a
$\approx50$ minute NML Cyg\ scan, followed by a $\approx$ 5 minute scan
of, alternately, the quasar J2015+3710 and J2012+4628.  For a
spectral-line observation, we observed a calibrator for about 3 minutes
and NML Cyg\ for 5 minutes.  Absolute flux density calibration was
established by an observation of 3C286 (J1331+3030), assuming a flux
density of 1.554 Jy at 43 GHz and 2.540 Jy at 22 GHz.  Data calibration
procedures are described in detail in \citet{1997ApJ...476..327R,
2007ApJ...671.2068R}.

\section{Results}

\subsection{Maser spectrum and spatial distribution}
\label{ssec:maser}

\begin{figure*}
  \centering
  \includegraphics[angle=0,scale=0.80]{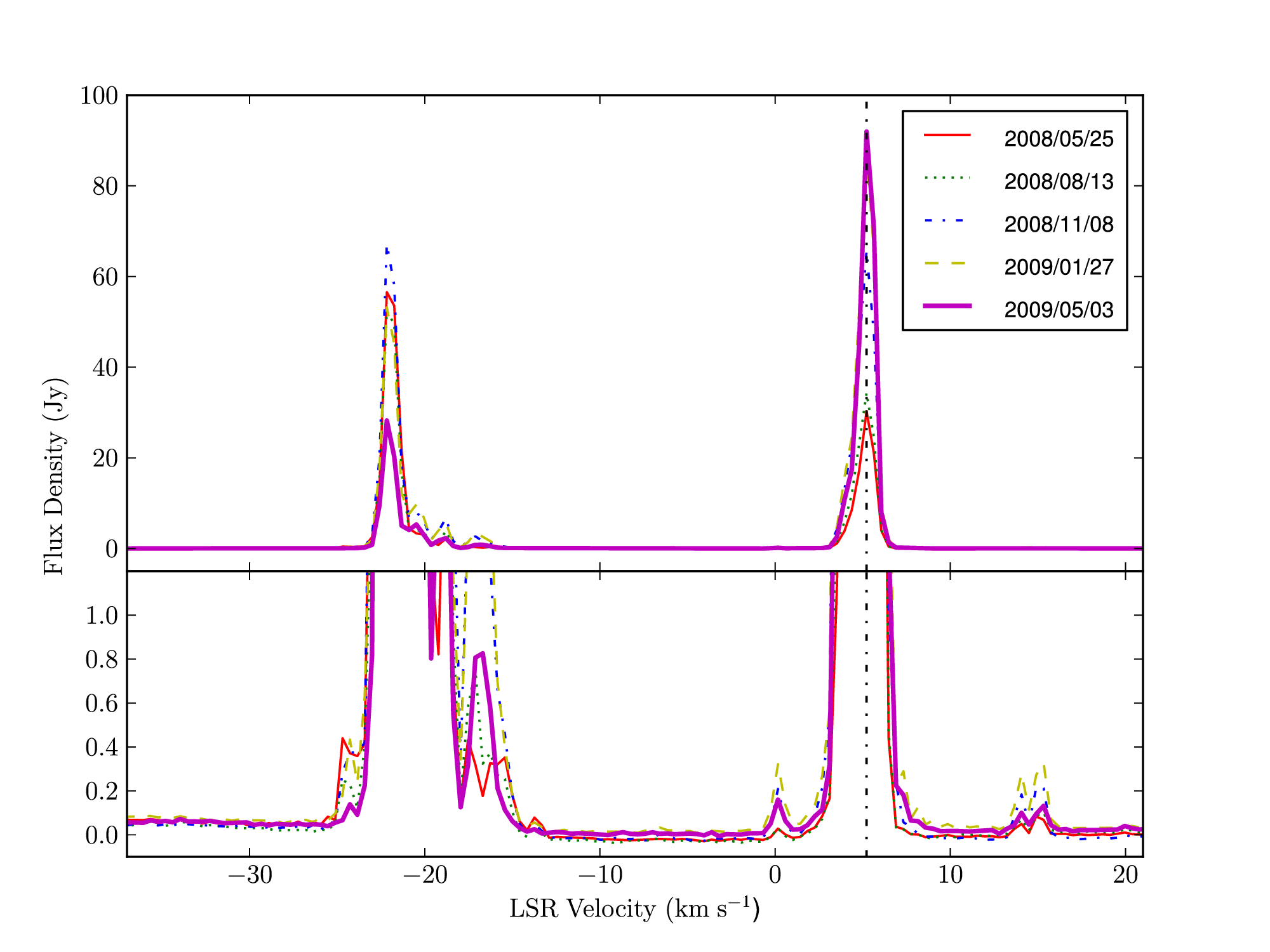}
  \caption{
Interferometer (scalar averaged cross-power amplitude over the full
duration of the observation) spectra of the \hho\ masers toward NML Cyg\
obtained with the inner 5 antennae of the VLBA at five epochs.  The
vertical {\it dash-doted line} indicates the maser feature at \VLSR\ of
5.22~\kms\ which served as the phase-reference. The {\it lower panel}
shows a 100-times blow-up of the \hho\ spectra shown in {\it upper
panel}.
\newline (A color version of this figure is available in the online journal.)
}
  \label{fig:nmlcyg_spec} 
\end{figure*}

Figure~\ref{fig:nmlcyg_spec} shows scalar averaged cross-power spectra
of \hho\ maser emission observed using only the inner 5 antennae of the
VLBA at all five epochs. The \hho\ maser emission spans a \VLSR\ range
from about $-$25 to 17~\kms, which is consistent with previous
observations \citep{1996MNRAS.282..665R,2008PASJ...60.1069N}.  One can
see that the flux densities of some maser features varied considerably
from epoch to epoch within about 1.0~yr, while the peaks of blue-shifted
(\VLSR\ = $-$22.17 \kms) and red-shifted (\VLSR\ = 5.22 \kms) features
appear more stable.  We confirmed that the brightest and most
blue-shifted component appears to exhibit a velocity drift, the LSR
velocities of this component in 1981, 1993, 2006 and 2008-2009 (this
paper) were $-$18, $-$20, $-$21.8 and $-$22.2 \kms, the resulting
acceleration is consistent with that derived by
\citet{2008PASJ...60.1077S}.

\begin{figure*}
  \centering
  \includegraphics[angle=0,scale=0.70]{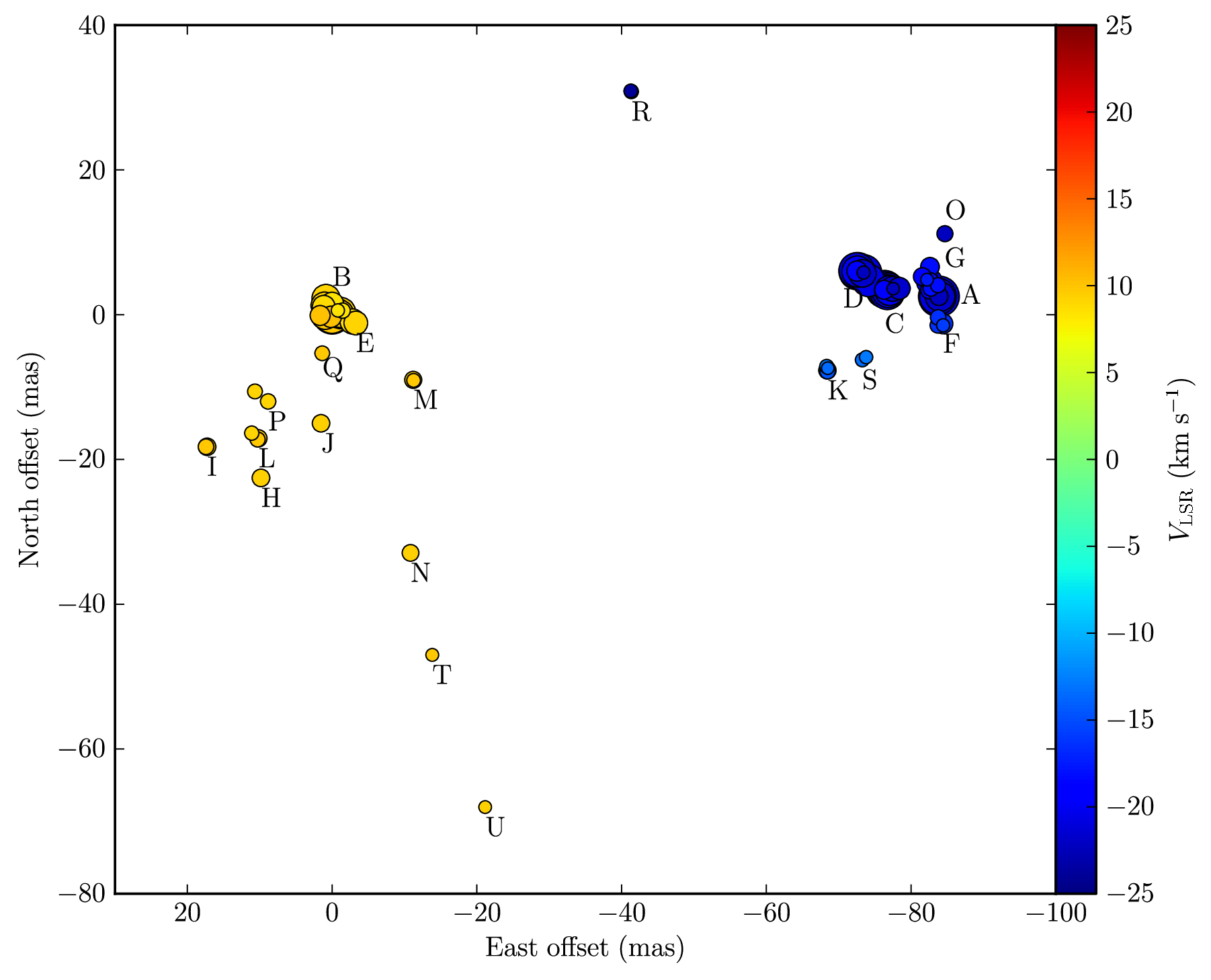}
  \caption{
Spatial distribution of the \hho\ maser features toward NML Cyg from
VLBA observations at five epochs.  Each maser spot is represented by a
{\it letter label} and a {\it filed circle} whose area is proportional
to the logarithm of the flux density, using the position and flux
density from the epoch at which it was first detected.  The color bar
denotes the \VLSR\ range from $-25$ to +25 \kms.  The reference maser spot
is in feature B, located at (0,0) mas.
\newline (A color version of this figure is available in the online journal.)
  \label{fig:h2o_maser_all}}
\end{figure*}

Figure~\ref{fig:h2o_maser_all} shows the spatial distribution of \hho\
maser emission toward NML Cyg\ relative to the reference maser spot at
\VLSR = 5.22~\kms\ from observations at five epochs.  We considered
maser spots at different epochs as being from the same feature if their
positions in the same spectral channel were coincident within $x$ mas
(where $x$ = 3 \masy\ $\times~\Delta t$ yr, and $\Delta t$ is the time
gap between two epochs), corresponding to a linear motion of less than
23 \kms.  Selecting the bright maser spots with flux density $>$ 0.5
\jybeam\ and SNR $>$ 10 in each channel, we found 21 features
that were detected at one or more epochs.

The total extent of the \hho\ maser spot distribution is about 120~mas,
which is consistent with previous studies from VLBI
observations~\citep{1996PhDT........82M, 2008PASJ...60.1069N}. Our first
epoch observation took place about 1 yr later than the last epoch in
\citet{2008PASJ...60.1069N}, and we found that the most red-shifted
maser cluster (including features B and E) and the most blue shifted
maser cluster (including features D and C) in
Fig.~\ref{fig:h2o_maser_all} are consistent with features at \VLSR\
about 6 and $-22$ \kms\ in \citet{2008PASJ...60.1069N}, respectively.
These two dominant maser clusters are about 70 mas apart. This suggests
that even though the spatial distribution changes in details, the maser
features at the \VLSR\ near the distribution of the masers in the two
peaks remains relatively stable.

\citet{1996MNRAS.282..665R} reported an irregular ring of emission
$\approx200$ mas across and a pair of outlying features to the
north-west (NW) and south-east (SE), nearly 600 mas apart, based on
MERLIN observations \citep{1994MNRAS.270..958Y,1996MNRAS.282..665R} and
VLA observations \citep{1985ApJ...290..660J}.  The prominent maser
features in the irregular ring are consistent with the features at
\VLSR\ about 5 and $-22$ \kms\ reported in this paper.  However, the NW
and SE maser features were not detected in our VLBA nor in previous VLBI
observations \citep{1996PhDT........82M,2008PASJ...60.1069N}.  For
our VLBI observations, typical diameters of the \hho\ maser spots are
$\approx$ 1 mas and the detected limitation is $\approx$ 10 mJy. This
corresponds to a brightness temperature lower limit of $\approx~4
\times~10^7$ K.  However, the brightness temperatures for the two
outlying features reported by \citet{1996MNRAS.282..665R} are
$\approx~10^7$ K. This might be the reason why the outlying features
were not detected with VLBI observations.

\begin{figure*}
  \centering
  \includegraphics[angle=0,scale=0.80]{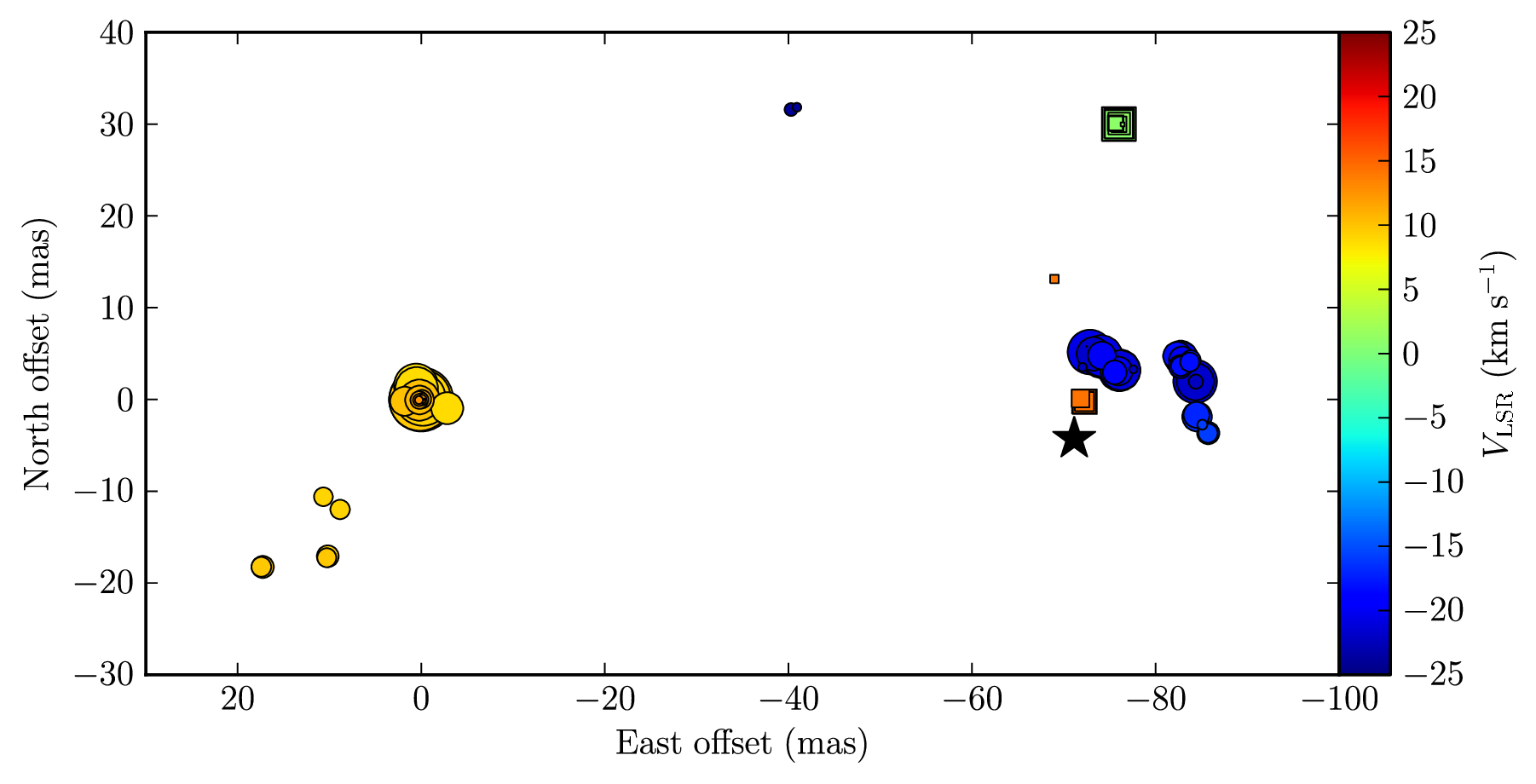}
  \caption{
Spatial distribution of the \hho\ ({\it circle }) maser and SiO ({\it
square}) maser spots toward NML Cyg\ from the fifth epoch observation.
Each maser spot is represented by a {\it filled circle or square} whose
area is proportional to the logarithm of the flux density. The
position of the central star is indicated by a {\it star} (see the text
in
\S~\ref{ssec:vla}).  The color
bar denotes the \VLSR\ range from $-25$ to 25 \kms\ of the maser features.
The reference \hho\ maser spot is located at (0,0) mas.
\newline (A color version of this figure is available in the online journal.)
  \label{fig:h2o_sio_maser}}
\end{figure*}

Unlike the \hho\ maser, the SiO maser emission toward NML Cyg\ during
our observation was very weak, and phase referencing with the maser
succeeded only at the last epoch.  Because we could register \hho\ and
SiO masers to the same background sources, the relative position between
the reference maser spots for two maser species should be accurate to
$\pm1$ mas (see Fig.~\ref{fig:h2o_sio_maser}).  The brightest SiO maser
features detected in our observations are at \VLSR\ about $-5.70$ and
11.03 \kms.  This differs from the VLBA observation of the same $v = 1,
J = 1 \to 0$ SiO maser emission by \citet{2000ApJ...545L.149B}.  The
extent of the maser distribution is about 30 mas and is consistent with
the diameter of the ring-like distribution from about 10 yr earlier.
>From a VLA observation of $v = 0, J = 1 \to 0$ SiO maser emission,
\citet{2004ApJ...608..480B} reported a radius of $\approx$ 50 mas for
the maser distribution, which is consistent with the inner boundary for
the dust shell at 2.13 $\mu$m~\citep{2001A&A...369..142B}.

\subsection{Parallax and proper motion}
\label{ssec:para}

We fitted elliptical Gaussian brightness distributions to the images of
strong maser spots and the extragalactic radio sources for all five
epochs.  The change in position of each maser spot relative to each
background radio source was modeled by the parallax sinusoid in both
coordinates (determined by a single parameter, the star's parallax) and
a linear proper motion in each coordinate.  The extragalactic source
with the best known position, \Jtwo, was relatively far (3.7\deg) from
NML Cyg, and it was only used to determine the absolute position of the
maser reference spot.  We used the other three calibrators, which are 
separated by less than 1.5\deg\ from NML Cyg, to determine the parallax.

\begin{figure*}
  \centering
    \includegraphics[angle=0,scale=0.80]{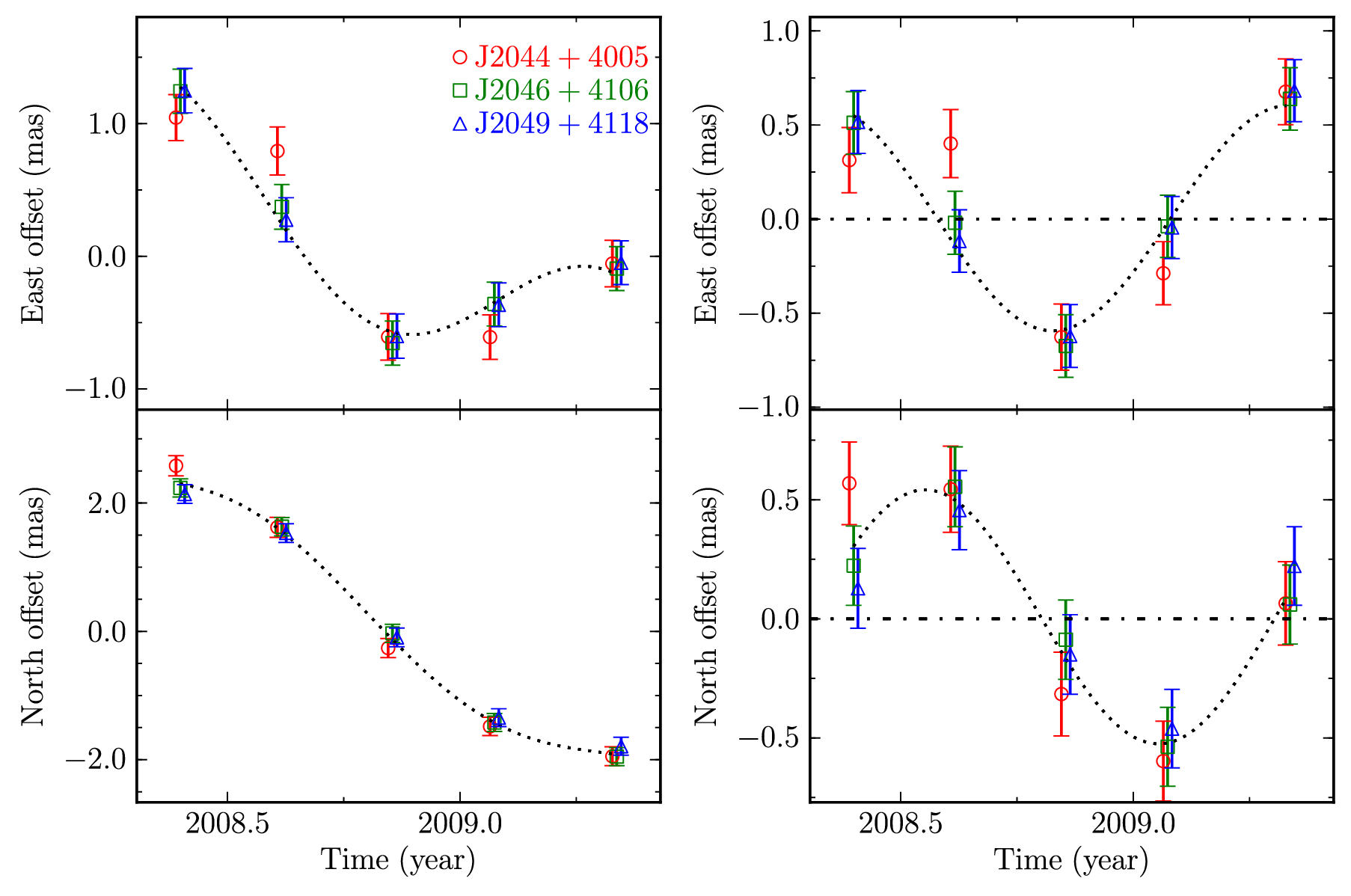}
    \caption{
Parallax and proper motion data ({\it markers}) and a
best-fitting model
({\it dotted lines}) that uses a combined estimate for the parallax and proper motion
for the maser spot at the \VLSR\ of 6.48~\kms.
Plotted are positions of the maser spot relative to the extragalactic
radio sources \Jfour\ ({\it circles}), \Jsix\ ({\it squares}) and
\Jnine\ ({\it triangles}).  {\it Left panel}: Eastward ({\it upper
panel}) and northward ({\it lower panel}) offsets versus time.  
{\it Right panel}: Same as the {\it left
panel}, except the best-fitting proper motion has been removed,
displaying only the parallax signature.
\newline (A color version of this figure is available in the online journal.)
    \label{fig:nmlcyg_para}}
\end{figure*}

As mentioned in \citet{2012ApJ...744...23Z}, the apparent motions of the
maser spots can be complicated by a combination of spectral blending and
changes in intensity.  Thus, for parallax fitting, one needs to find
stable, unblended spots and/or use many maser spots to average out these
effects.  Thus, we first fitted a parallax and proper motion to the
position offsets for each \hho\ maser spot relative to each background
source separately.  In Fig.~\ref{fig:nmlcyg_para}, we plot the position
of the maser spot at \VLSR\ = 6.48 \kms\ (the first maser spot listed in
Table~\ref{tab:para_pm}) relative to the three background radio sources
as an example, with superposed curves representing a model that uses a
combined estimate for the parallax and proper motion as described below.
This is one of the most red-shifted compact and unblended spots that was
detectable at all five epochs.

Table~\ref{tab:para_pm} shows the independently estimated parallaxes and
proper motions for each maser spot and extragalactic source pair.  While
the parallaxes should be identical within measurement uncertainties, the
proper motions are expected to vary among maser spots owing to internal
motions of $\approx15$~\kms\ ($1.9$ \masy\ at a distance of 1.6 kpc).
While most of the maser spot parallaxes show good internal consistency,
the dispersion of parallax estimates over the entire ensemble is
considerably larger than the formal errors would suggest.  This is
caused by residual systematic errors affecting the fits, which originate
in the complexity and evolution of blended spectral and spatial
structure for some of the masers.

Among these independent solutions, we found the parallaxes associated
with maser spot at \VLSR\ = $-19.64$ \kms\ had the largest relative
uncertainties of $\approx$ 200~$\mu$as.  The spots in this feature
are likely affected by line blended, and we discarded the data for
this feature. We used all remaining position data in a combined
solution, which used a single parallax parameter for all
maser spots and extragalactic source pairs (but allowed for different
proper motions for different maser spots).

The combined parallax estimate is $0.620 \pm 0.047$~mas, corresponding
to a distance of $1.61^{+0.13}_{-0.11}$~kpc, placing NML Cyg possibly in
a Local (Orion) arm.  The quoted uncertainty is the formal error
multiplied by $\sqrt{n}$ (where $n = 18$ is the number of maser spots
used in the final parallax fit) to allow for the possibility of
correlated position variations for all the maser spots.  This could
result from small variations in the background source or from un-modeled
atmospheric delays, both of which would affect the maser spots nearly
identically~\citep{2009ApJ...693..397R}.  The un-weighted average
absolute proper motion of all selected maser spots is \mux\ = $-1.55\pm
0.42$ \masy\ and \muy\ = $-4.59 \pm 0.41$ \masy, where \mux\ = 
$\mu_{\alpha}\cos{\delta}$ and \muy\ = $\mu_{\delta}$.

\addtocounter{table}{1}


Combining the parallax and proper motion measurements with the systemic
\VLSR\ of $-1.0 \pm 2$ \kms\
\citep{2003A&A...407..609K,2004MNRAS.348...34E} enable us to determine
the three-dimensional peculiar motion (relative to circular motion
around the Galactic center) of NML Cyg.  Adopting a flat rotation curve
for the Milky Way with rotation speed of LSR \To\ = $239 \pm 7$ \kms,
distance to the Galactic center of \Ro\ = $8.3 \pm 0.23$ kpc
\citep{2011AN....332..461B}, and the solar motion of
(\U=$11.1^{+0.69}_{-0.75}$, \V=$12.24^{+0.47}_{-0.47}$,
\W=$^{+0.37}_{-0.36}$) \kms\ from Hipparcos measurement revised by
\citet{2010MNRAS.403.1829S}, we estimate a peculiar motion for NML Cyg
of (\Us=$-2.5 \pm 3.8$, \Vs=$-6.1 \pm 2.1$, \Ws=$-4.8 \pm 3.3$) \kms,
where \Us, \Vs, \Ws\ are directed toward the Galactic center, in the
direction of Galactic rotation and toward the North Galactic Pole (NGP),
respectively.  Thus, NML Cyg is orbiting the Galaxy slower than
expected from a circular orbit for a flat rotation curve, similarly to
the average of many massive star forming regions found by
\citet{2009ApJ...700..137R}.  The peculiar velocity components toward
the Galactic center and the NGP are relatively small with somewhat
larger uncertainties.

\subsection{Radio photosphere and circumstellar maser emission from VLA
observation}
\label{ssec:vla}

\begin{figure}
  \begin{center} \includegraphics[scale=0.5]{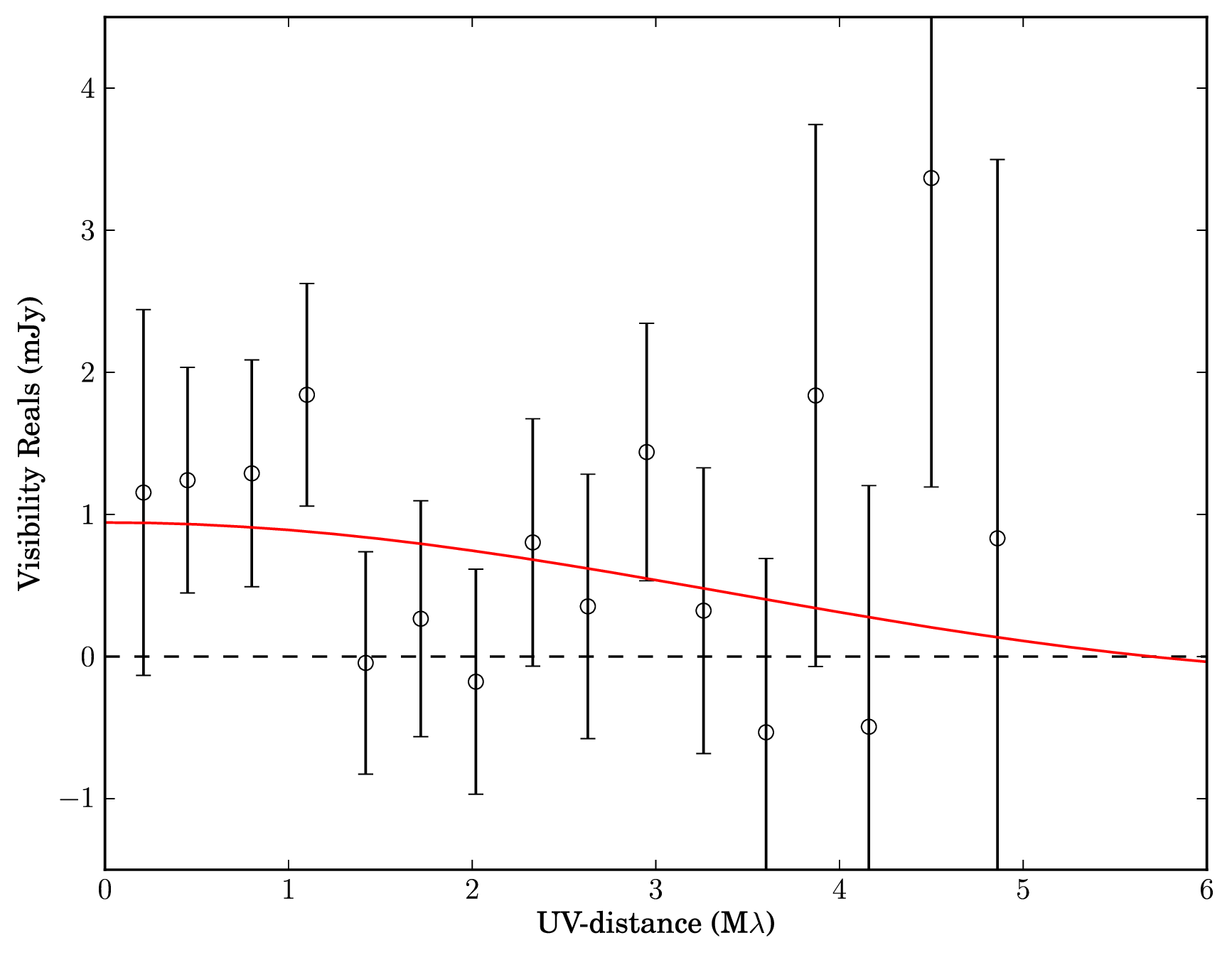}
  \end{center} \caption{
Fringe visibility vs. baseline length measured for NML Cyg at 43 GHz on
2008 December 20 under VLA program AZ178.  The {\it solid line}
indicates a uniformly bright circular disk model with diameter of 44 mas.
  } \label{fig:vis_fit} \end{figure}

\begin{figure*}
  \begin{center}
    \includegraphics[scale=0.7]{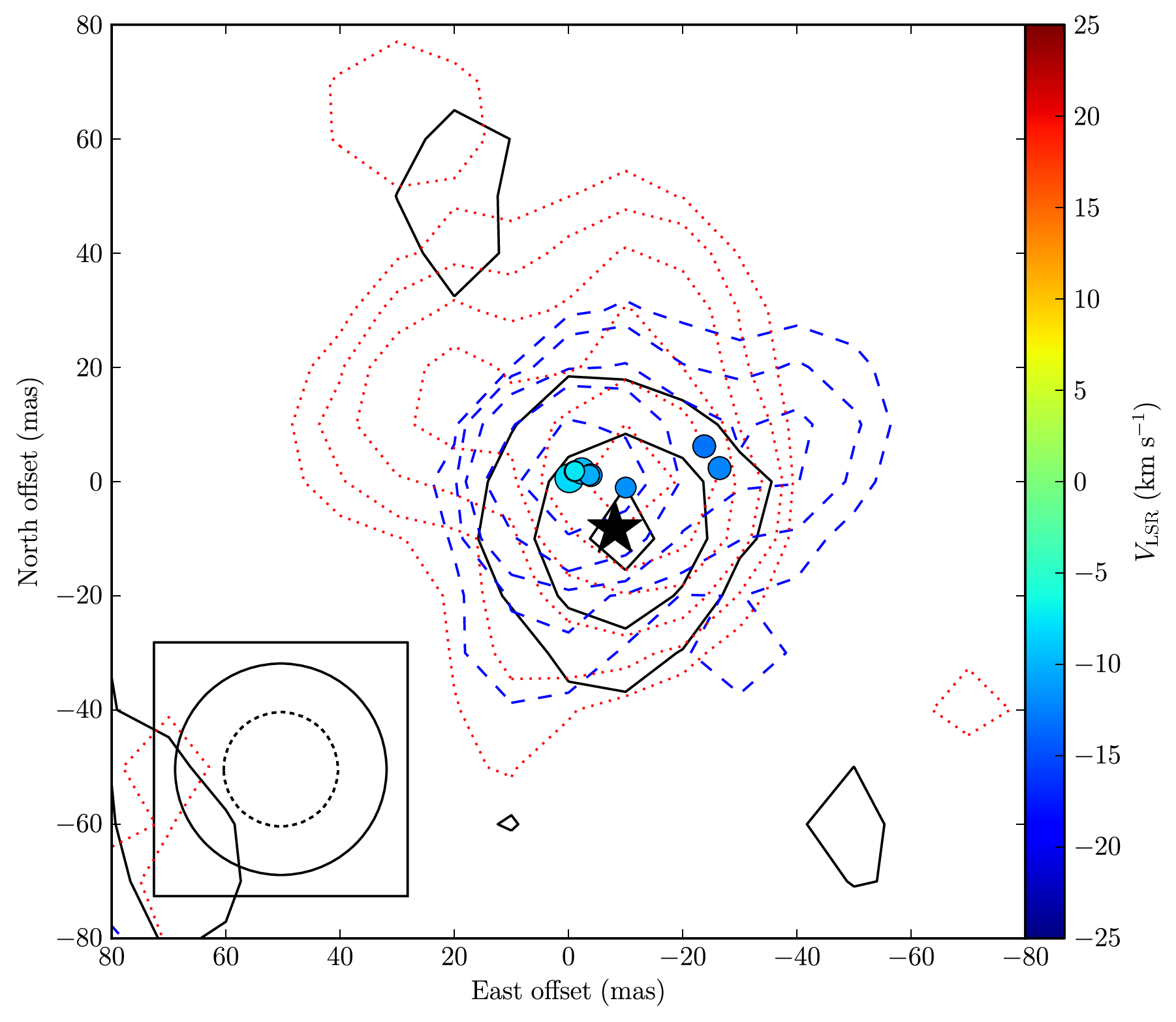}
  \end{center}
  \caption{
  SiO maser emissions integrated over \VLSR\ from $-28.86$ to $-7.14$
  \kms\ ({\it dashed contour}) which is within the same \VLSR\ range as
  that of the narrow-band data, and \VLSR\ from $-7.14$ to 22.93 \kms\
  ({\it dotted contour}) toward NML Cyg\ superposed on the 43 GHz
  continuum emission ({\it solid contour}) with peak position indicated
  by the {\it star}. The maser spots within the same \VLSR\ range as the
  narrow-band data are marked with circles, with size proportional to
  the logarithm of their flux densities. The \VLSR\ of the maser spots
  is color-coded as indicated by the color bar on the right side, which
  use the same range of \VLSR\ as other maser spots figures in this
  paper for comparison.  All offsets are relative to the strongest maser
  emission at \VLSR\ of $-4.66$ \kms.  The relative positions of the SiO
  maser and radio continuum emission is estimated to be accurate to
  $\approx 3$ mas.  Contour levels for integrated maser emission are
  0.18 \jybeam\ $\times~2^n, n = 0 \cdots 5$ and for the radio
  continuum are integer multiples of 0.10 \mjybeam.  The restoring beam
  for the maser ({\it dotted circle}) and continuum ({\it solid circle})
  emissions are indicated in the lower left corner.
\newline (A color version of this figure is available in the online journal.)
  }
  \label{fig:rp_sio}
\end{figure*}

We detected the radio photosphere of NML Cyg at 43 GHz and made a
uniformly weighted image (see Fig.~\ref{fig:rp_sio}).  We also fitted a
model of an uniformly bright disk to the $uv$-data directly using the
AIPS task OMFIT, obtained a diameter of $44 \pm 16$ mas and integrated
flux density of $0.82 \pm 0.11$ mJy. We binned the data and plotted the
real part of the visibility versus baseline length in
Fig.~\ref{fig:vis_fit}.

Taking the crude characteristics of the radio photosphere to be \Rrp\
$\approx 2~\times$ \Rop, and \Trp\ $\approx 0.71~\times$ \Top, which are
inferred from the imaging of variable stars by
\citet{1997ApJ...476..327R} (where \Rrp\ and \Rop\ are the radii of the
radio and optical photospheres, respectively, and \Trp\ and \Top\ are the
effective temperatures of the radio and optical photospheres), the 
semi-diameter of radio photosphere, 22 mas, is reasonably consistent with
that of 16.2 mas derived by \citet{2001A&A...369..142B}.

\begin{figure*}
  \begin{center}
    \includegraphics[scale=0.7]{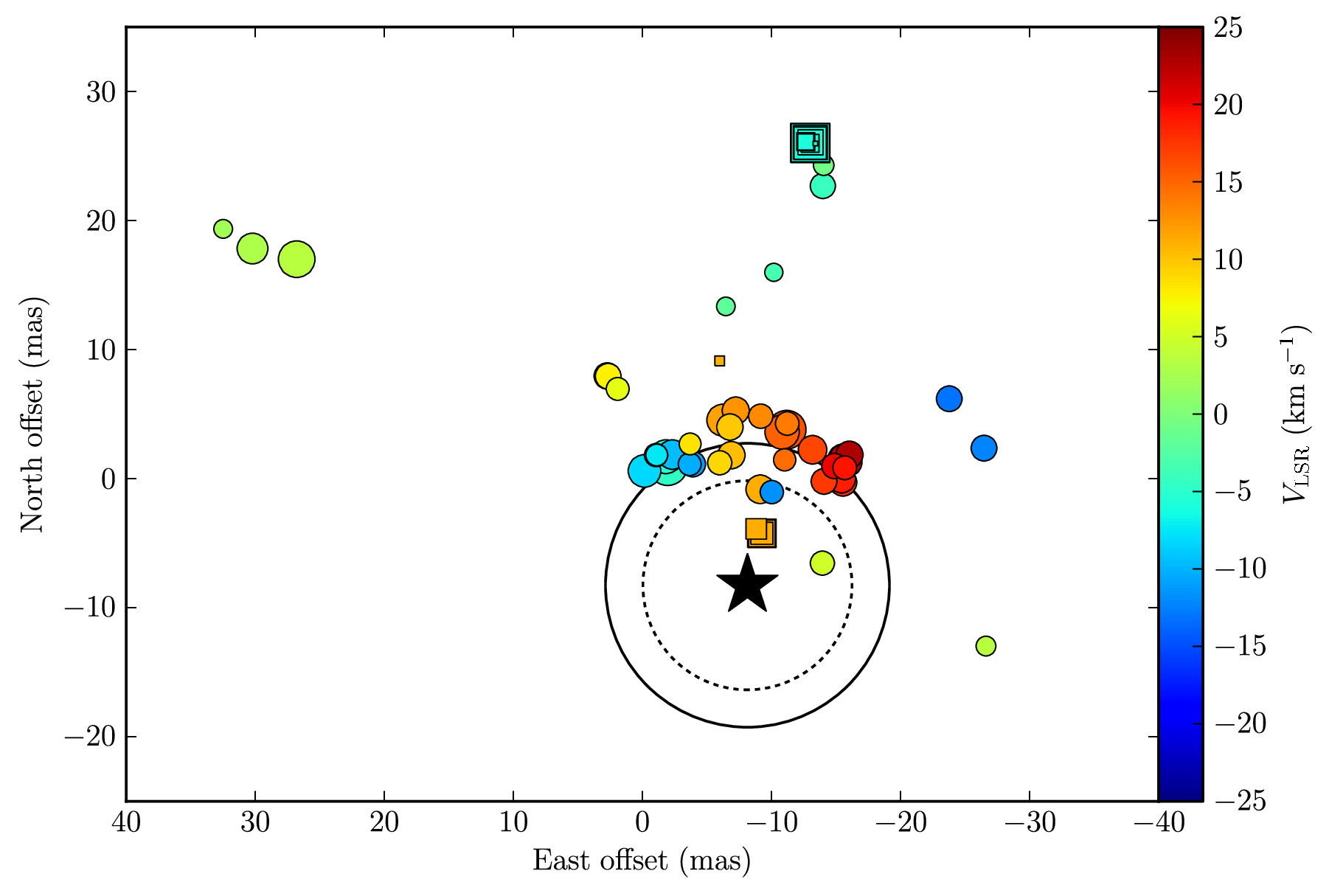}
  \end{center}
  \caption{
Comparison of SiO maser distribution from the VLBA observation ({\it squares}) on
2009 May 25 and VLA observation ({\it circles}) on 2008 December 20, after
cross-registration (see the text). The sizes of the markers are
proportional to the logarithm of their flux densities.  The \VLSR\ of the
maser spots is color-coded as indicated by the color bar on the right
side.  The position of the central star is indicated as the {\it
star}. The {\it solid circle} and {\it dotted circle} denote
the stellar diameter of 22 mas from this paper and 16 mas from
\citet{2001A&A...369..142B}, respectively.
\newline (A color version of this figure is available in the online journal.)
  }
  \label{fig:vlba_vla_sio}
\end{figure*}

\begin{figure*}
  \begin{center}
    \includegraphics[scale=0.7]{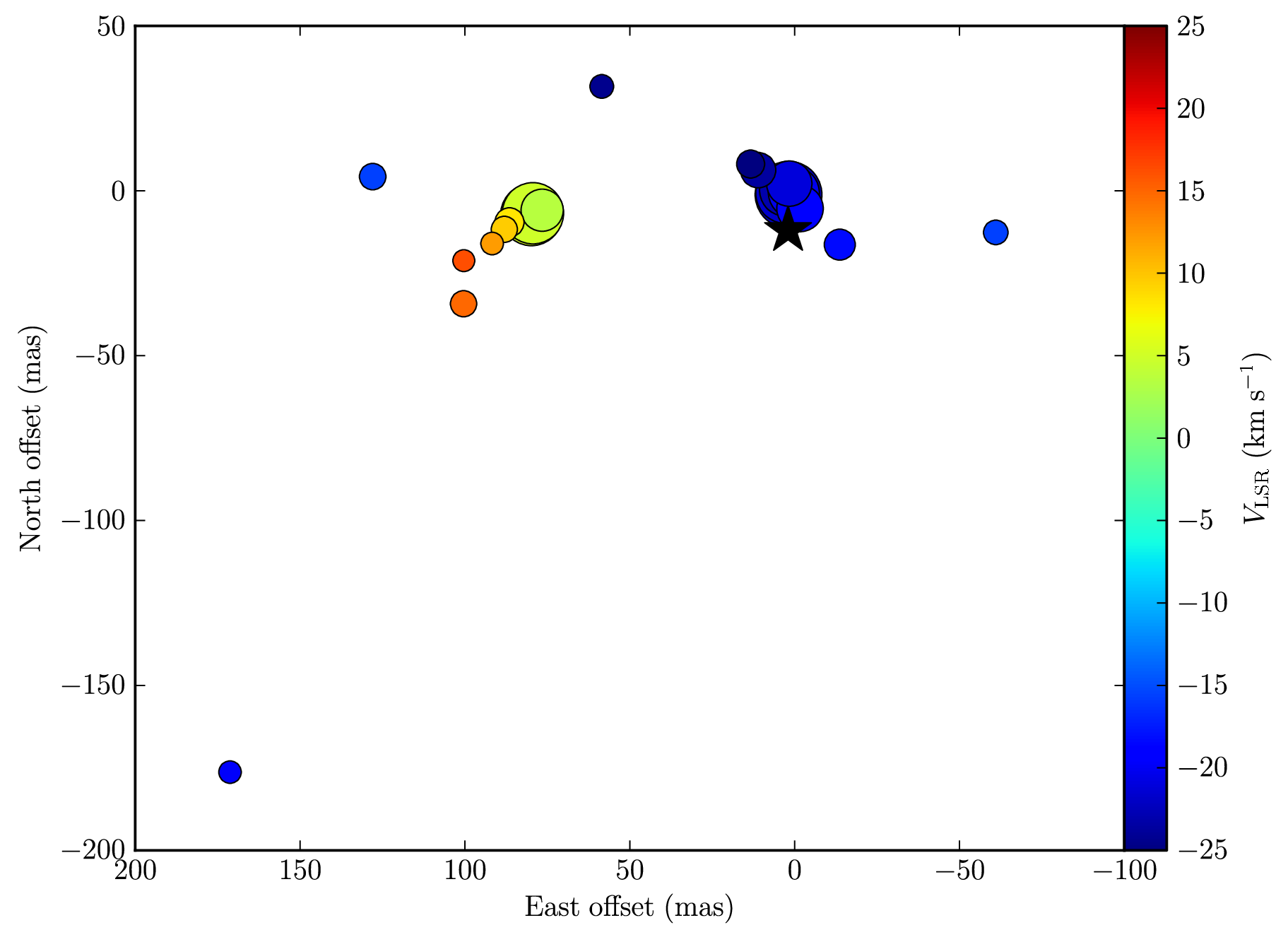}
  \end{center}
  \caption{
  \hho\ maser emissions distribution toward NML Cyg from VLA observation
  on 2008 December 20.  All offsets are relative to the strongest maser
  emission at \VLSR\ of $-22.25$ \kms. All the maser spots are marked
  with {\it circles}, with sizes proportional to the logarithm of their flux
  densities.  The
  position of the central star is indicated by a {\it star}.
\newline (A color version of this figure is available in the online journal.)
  }
  \label{fig:vla_h2o}
\end{figure*}

In order to register the SiO masers relative to the radio photosphere,
we first measured the position of the radio photosphere from the
broad-band data relative the maser emission from the narrow-band data.
Next, we aligned the maser emission from the spectral-line data by
producing a map of pseudo-continuum data by integrating the line data
over the entire narrow-band. Because the narrow-band and the
pseudo-continuum data cover the maser emission from the same velocity
range, we can align them by comparing the positions of peaks in each
map. This allows the positions of the emission in individual channels to
be registered to the map of narrow-band data, and then to the radio
photosphere. We estimated a position offset of the weak continuum source
(radio photosphere) relative to the reference maser spot to be (\dx\ =
$-8$, \dy\ = $-8$) mas with an uncertainty of $\approx$ 3 mas.
Fig.~\ref{fig:rp_sio} shows the relative positions of the SiO maser and
43 GHz continuum emission used to register the images.

The relative offsets of the peaks of SiO maser emission and the central
star agree well with observations of Mira variable stars
\citep{2007ApJ...671.2068R}, which show the SiO maser emission surrounds
the central star in a partial ring with radii of 2 -- 3 \Rop.  For NML
Cyg the SiO radius is also consistent with the dust inner-radius of 45
mas \citep{1986ApJ...302..662R}.

In order to estimate the absolute position of the radio photosphere
(which should be that of the central star), we have to align the SiO
maser spots measured with the VLA with those from the VLBA observations.
As discussed in \citet{2012ApJ...744...23Z},  if a single spectral
channel contains emission from across the source, the VLA position would
be falsely interpreted as being near the center of the distribution.
Furthermore, the VLBA observing date of SiO emission is about five
months later than that of VLA. So, there is no guarantee that the VLA
and VLBA maps should be identical.  However, we found that the two maser
clusters at \VLSR\ about 11 and $-6$ \kms\ in the VLBA map are in good
agreement with those at similar \VLSR\ values in the VLA map, after
shifted the VLBA map (\dx\ = $-13$, \dy\ = 26) mas, with an uncertainty
of $\approx5$ mas.  Thus, we can register the radio photosphere to the
VLBA SiO maser spots with a conservative uncertainty of 10 mas (see
Fig.~\ref{fig:vlba_vla_sio}).

The absolute position of the reference SiO maser spot can be estimated
relative to the extragalactic sources, yielding the absolute stellar
position of NML Cyg at epoch 2008.868 of \ra\ = 20\h 46\m 25\decs5382
$\pm$ 0\decs0010, \dec\ = 40\deg 06\arcmin59\decas379 $\pm$ 0\decas015.
The absolute position is consistent with, but considerably more accurate
than, that from the Two Micron All Sky Survey (2MASS) by
\citet{2006AJ....131.1163S} (after converting to J2000 coordinates and
accounting for our measured absolute proper motion); differences of 39
mas in right ascension and 10 mas in declination are well within the 170
mas uncertainty of the 2MASS position.

Fig.~\ref{fig:vla_h2o} shows the \hho\ maser distribution from our VLA
observation. Similar to the map as shown in
Fig.~\ref{fig:h2o_maser_all}, the maser distribution is dominated by two
maser clusters with a separation of $\approx$ 70 mas.  However, the
total extent of the maser distribution is much larger than that of VLBA
map. The outlying maser spot to the south-east in the VLA map is likely
associated with a feature reported by \citet{1996MNRAS.282..665R}.

\section{Discussion}

\subsection{Relation to the Cyg region and OB association}

\citet{1982A&A...108..412H} observed a \HII\ region located
northwest of NML Cyg, and \citet{1983ApJ...267..179M} suggested that
it represented a portion of NML Cyg's circumstellar envelope
photoionized by the luminous and hot stars in the Cyg OB2 association.
Cyg OB2 lies near the center of the X-ray emitting Cygnus X super
bubble, which is composed of numerous individual \HII\ regions, a number
of Wolf-Rayet and O3 stars and several OB associations.

\begin{figure*}
  \centering
  \includegraphics[angle=0,scale=0.70]{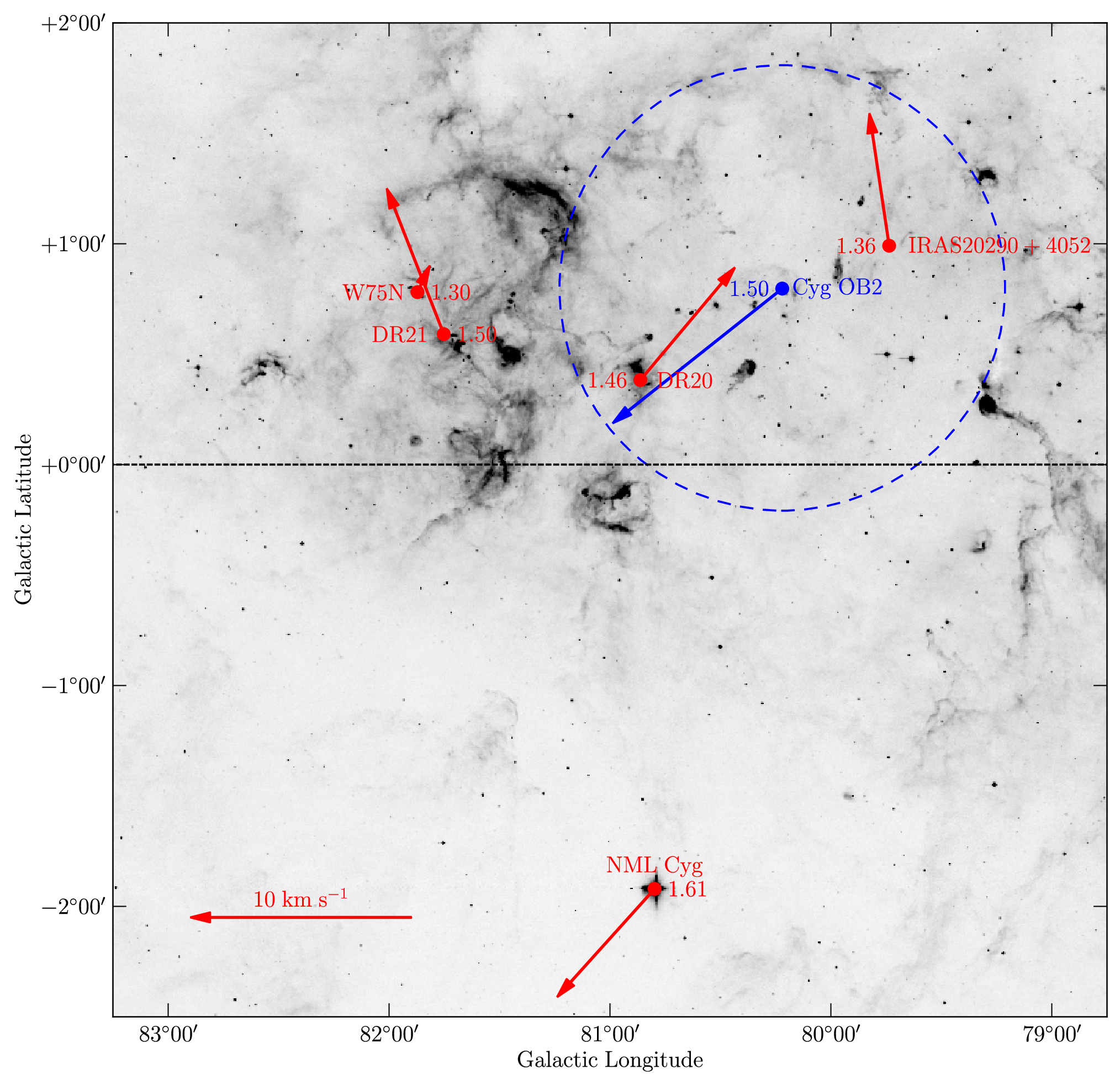}
  \caption{
Midcourse Space eXperiment (MSX) 8 $\mu$m image of the Cygnus X region
overlaid with the peculiar motions for maser sources with parallax and
proper motion measurements.  The {\it dotted line} indicates the Galactic
plane.  The {\it dots} mark the water maser (NML Cyg, this paper) and
the methanol maser sources \citep{2012A&A...539A..79R} and the center of
Cyg OB2 \citep{2005A&A...438.1163K}, with their distances labeled in
kpc.  The peculiar motions of sources are indicated with arrows.  The
approximate size of the Cyg OB2 association is denoted by a {\it circle}
with diameter of 2\deg\ \citep{2000A&A...360..539K}.
\newline (A color version of this figure is available in the online journal.)
  \label{fig:cyg-x}}
\end{figure*}

Recently, \citet{2012A&A...539A..79R} reported distances from
trigonometric parallax measurements of four 6.7 GHz methanol maser
sources in the Cygnus X region that are consistent with $1.40 \pm 0.08$
kpc for this region (see Fig.~\ref{fig:cyg-x}).  This is similar to the
distance of 1.45 kpc to Cyg OB2 obtained by \citet{2003ApJ...597..957H},
based on fitting 35 dwarf cluster members with spectral types between
O7.5 and B1 to a $2 \times 10^6$ yr isochrone, and also consistent with
latest estimated photometric distance of 1.5 kpc determined by
\citet{2005A&A...438.1163K}, although slightly smaller than previous
work quoting 1.74 kpc \citep{1991AJ....101.1408M}.

Given that the Cygnus X region contains huge amounts of dust, one might
expect that this dust would extinguish the optical light from Cyg OB2.
However, little extinction toward Cyg OB2 was found from the NIR
extinction map of the region \citep{2007A&A...476.1243M}.  We also note
that the interstellar medium (ISM) absorption signatures from NML Cyg,
seen in Herschel spectra of the \hho\ ortho ($1_{1,0} \to 1_{0,1}$) and
para ground-state ($1_{1,0} \to 0_{0,0}$) line
\citep{2012InPreparation.T} and low$-J$ CO transitions
\citep{2003A&A...407..609K}, suggest that they arise from a cold gas
component. One possible explanation is that the diffuse gas in Cygnus X
is located in front of NML Cyg.  Taking our distance of
$1.61^{+0.13}_{-0.11}$ kpc to NML Cyg, we offer new observational
evidence that NML Cyg might be associated with Cyg OB2.

\begin{figure*}
  \centering
  \includegraphics[angle=0,scale=0.80]{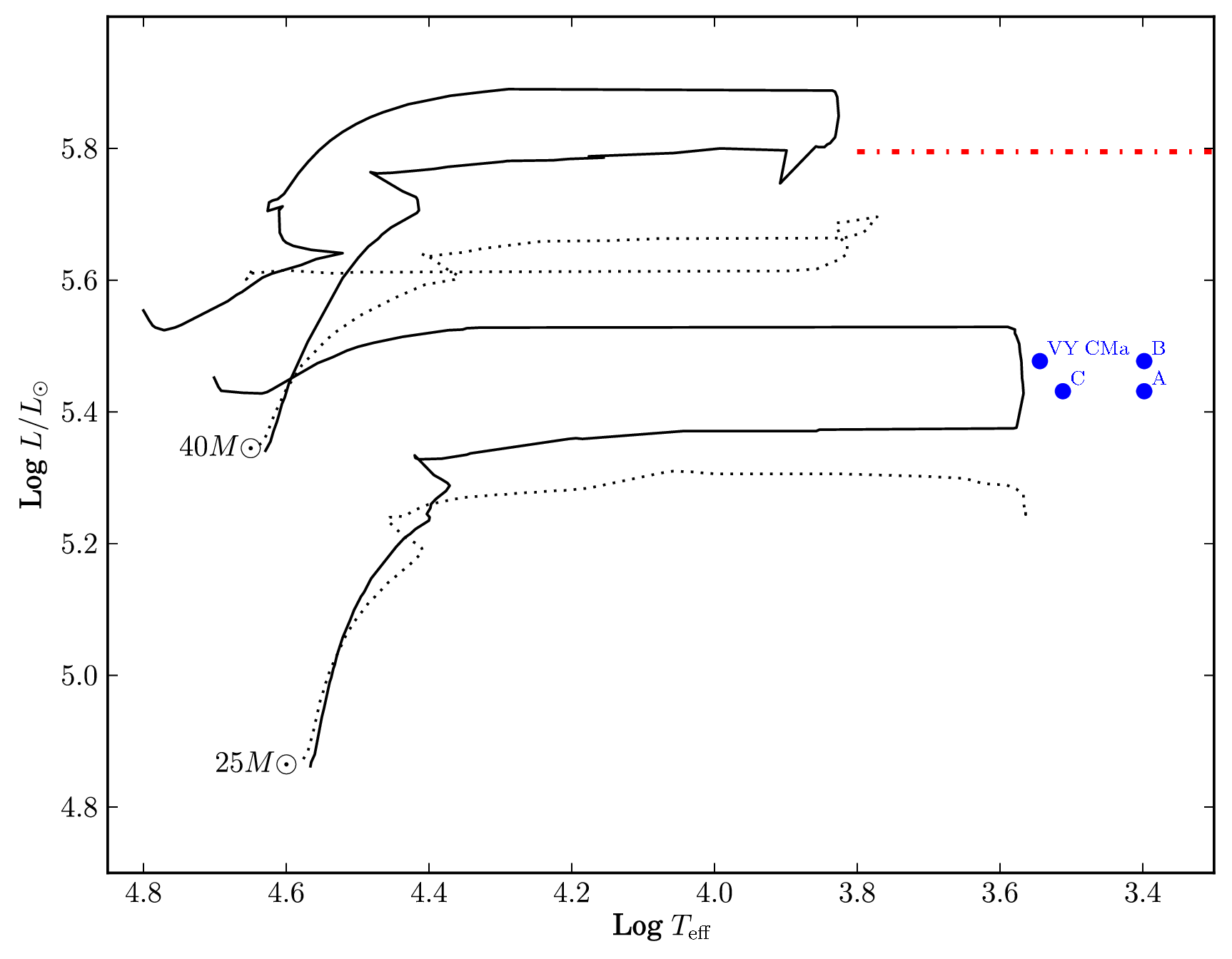}
  \caption{
Evolutionary tracks of 25 and 40 \msun\ non-rotating ({\it dotted
lines}) and rotating ({\it solid lines}) with a velocity of
300 \kms.  The model assumes a star with near solar metallicity ($Z = 0.02$)
evolving from the zero-age main sequence to the red supergiant and then
Wolf-Rayet star phase.  The dot-dashed line denotes the empirical upper
luminosity boundary (for \Teff\ less than 6000 K) from
\citet{2006AJ....131..603S}.  The {\it filled circle} labeled ``VY CMa''
represents the results for that star from \citet{2008PASJ...60.1007C}.
For NML Cyg the {\it filled circle} labeled ``A'' corresponds to \Teff
= 2500 K
\citep{1983MNRAS.202..767R,1997ApJ...481..420M,2004ApJ...610..427Z} and
$2.7 \times 10^5$ \Lsun\ revised with the distance in this paper; {\it
filled circles} ``B''  and ``C'' are for \Teff = 2500 K and $3.0 \times
10^5$ \Lsun\ \citep{2009ApJ...699.1423S} and \Teff = 3250 K
\citep{1986ApJ...302..662R} and $2.7 \times 10^5$ \Lsun. 
  \label{fig:evo}}
\end{figure*}

\begin{table*}
 \caption{\label{tab:comp} Comparison of astrometric parameters of
 red hypergiants and its associated clusters}
 \begin{center}
 \begin{tabular}{lccccccccc}
 \hline \hline
 \tabincell{c}{Star    \\          \\ (1)}  &
 \tabincell{c}{Cluster \\   \\ (2)} &
 \tabincell{c}{Separation \\ (\deg)  \\ (3)} &
 \tabincell{c}{Distance     \\ (pc) \\ (4)} &
 \tabincell{c}{\mux\  \\ (\masy)  \\ (5)} &
 \tabincell{c}{\muy\  \\ (\masy)  \\ (6)} &
 \tabincell{c}{\VLSR\  \\ (\kms)  \\ (7)} & \\
 \hline
NML Cyg  &         &      &  1610 $\pm$ 120 & $ -1.55 \pm 0.42$ & $-4.59 \pm 0.41$ & $-01.0$ \\
         & Cyg OB2 & 2.7  &  1500           & $ -1.22 \pm 0.20$ & $-4.72 \pm 0.24$ & $+07.0 \pm 6.0$\\
         &&& &&& & \\
VY CMa   &         &      &  1200 $\pm$ 120 & $ -2.80 \pm 0.20$ & $+2.60 \pm 0.20$ & $+22.0$ \\
         & NGC2362 & 1.3  &  1389           & $ -2.03 \pm 0.31$ & $+2.47 \pm 0.34$ & $+06.2 \pm 8.9$\\
         &&& &&& & \\
S Per    &         &      &  2400 $\pm$ 100 & $ -0.49 \pm 0.23$ & $-1.19 \pm 0.20$ & $-38.5$ \\
         & h Per   & 1.5  &  2079           & $ -0.27 \pm 0.26$ & $-0.90 \pm 0.39$ & $-35.7 \pm 2.0$\\
         &$\chi$ Per& 1.5 &  2345           & $ -0.46 \pm 0.24$ & $-0.23 \pm 0.50$ & $-35.8 \pm 7.1$\\
 \hline
 \end{tabular}
 \end{center}
 \tablefoot{The first and second columns list the names of a red hypergiant
 and its associated cluster. The third column lists the separation between
 a star and the center of the cluster. The fourth to seventh columns list
 distance, proper motion eastward and northward and LSR velocity of red
 hypergiant or cluster. Parameters for NML Cyg are from this paper;
 parameters for VY CMa are from \citet{2012ApJ...744...23Z}, except that
 \VLSR\ is from \citet{2006A&A...454L.107M}; Parameters for S Per are
 from \citet{2010ApJ...721..267A}.  Parameters for clusters are from
 \citet{2005A&A...438.1163K}, except that \VLSR\ of Cyg OB2 is from
 \citet{2007ApJ...664.1102K, 2008ApJ...681..735K}.}
\end{table*}

As listed in Table~\ref{tab:comp}, our measured absolute proper motion
of NML Cyg is in good agreement with that of Cyg OB2, computed by
averaging proper motions of the most probable cluster members
\citep{2005A&A...438.1163K}.  NML Cyg's \VLSR\ of $-1 \pm 2$ \kms\ is consist
with the systemic \VLSR\ of 7 \kms\ (helio-centric velocity $-10.3$
\kms), within a dispersion of 6 \kms, for Cyg OB2 as estimated by
\citet{2007ApJ...664.1102K, 2008ApJ...681..735K}.  For comparison,
Table~\ref{tab:comp} also lists parameters for two other red hypergiants
and their associated clusters. We find that the radio proper motions of
stars and the optical proper motion of their associated clusters are in
good agreement, providing new observational evidence for their
association.

%
%
Figure~\ref{fig:cyg-x} shows the positions of maser sources in the
Cygnus X region with accurate parallax measurements and their peculiar
motions projected on the sky.  We find that the methanol sources are
moving upward (toward increasing Galactic latitude) while NML Cyg and
Cyg OB2 are moving downward from the Galactic plane.  This suggest that
the space motions in the Cygnus X region are much more complicated than
expected from an expanding Str\"omgren sphere centered on Cyg OB2 with
an angular diameter of 4\deg\ \citep{2003IAUS..212..505K}.

\subsection{Stellar parameters}


\citet{2009ApJ...699.1423S} determined NML Cyg's minimum bolometric
luminosity, \Lbol\ = $(1.04 \pm 0.05) \times 10^5 \cdot $ \dLsun, by
integrating its SED from 0.5 to 100 $\mu$m without correction for
extinction.  This luminosity is in good agreement with the estimate of
\Lbol\ = $1.13 \times 10^5 \cdot $ \dLsun\ by
\citet{2001A&A...369..142B}. Adopting our measured distance of
$1.61^{+0.13}_{-0.11}$ kpc, we find NML Cyg's luminosity to be $(2.7 \pm
0.5) \times 10^5$ \Lsun.
%
We note that a stellar temperature of 2500 K, first derived from a model
by \citet{1983MNRAS.202..767R}, gives a good fit to the infrared data
\citep{1997ApJ...481..420M}, as compared to a temperature of 3250 K
obtained from spectroscopy by \citet{1986ApJ...302..662R}.

To examine the evolutionary status of NML Cyg, we compared its position
on the H-R diagram with stellar evolutionary tracks
\citep{2003A&A...404..975M}, using different \Teff\ and luminosity
values (see Fig.~\ref{fig:evo}).  Although NML Cyg's position with
\Teff\ of 3250 K is more consistent with the evolutionary track than
those with \Teff\ of 2500 K (compare dots `C' with `A'), we find that
NML Cyg's position is closest to the evolutionary track with an initial
mass of 25 \msun, instead of 40 \msun\ quoted in
\citet{2009ApJ...699.1423S}, no matter which luminosity and \Teff\ we
adopted (see dots `A', `B' and `C' in Fig.~\ref{fig:evo}).  This mass is
similar to that of VY CMa, another red hypergaint star, estimated by
\citet{2008PASJ...60.1007C}.  Re-scaling NML Cyg's luminosity (compare
dots `A' with `B') with our distance drops it in the H-R diagram well
below the empirical upper luminosity boundary (see the red dot-dashed
line in Fig.~\ref{fig:evo}), which is not sensitive to temperature for
stars $< 6000$ K.  However, NML Cyg's position on the H-R diagram,
compared with the evolutionary track, constrains its age (by adopting
the evolutionary time of the closest point to NML Cyg on the track) to
be $\approx$ 8 Myr, which is considerably greater than that of $\approx$
2 -- 3 Myr for Cyg OB2 \citep{2001AJ....121.1050M, 2003ApJ...597..957H}.
Since the observational evidence now confirms that NML Cyg is
associated with Cyg OB2, they are probably about the same age. If so,
the stellar evolutionary model cannot be appropriate for NML Cyg.

NML Cyg's stellar size of 16.2 mas from \citet{2001A&A...369..142B} was
derived using the Stepan-Boltzmann law, adopting \Teff\ = 2500 K and a
distance of 1.74 kpc.  Rescaling this stellar diameter with our distance
of 1.61 kpc gives 15.0 mas.  For the radio photosphere, we fitted a
round-disk model with a diameter of $44 \pm 16$ mas and flux density of
$0.8 \pm 0.1$ mJy using the AIPS task OMFIT,
and we calculate a (Rayleigh-Jeans) \Teff\ between 100 K and 6000 K
(i.e., $\pm 2\sigma$).  Clearly more observations are needed to improve
this temperature constraint.

\subsection{Maser distribution relative to radio photosphere}

\citet{1996MNRAS.282..665R} proposed a bipolar \hho\ maser outflow and
suggested the central star is located close to the prominent \hho\ maser
cluster at \VLSR\ about $-22$ \kms\ in the irregular ring, assuming the
NW and SE feature lie approximately symmetrically about the star.  Based
on high-resolution HST observations, \citet{2006AJ....131..603S} find
that the circumstellar material of NML Cyg shows an ``arc-like'' shape
with a line of symmetry and an axis facing toward Cyg OB2, which is
coincident with the \hho\ maser \citep{1996MNRAS.282..665R} and ground
state SiO distribution \citep{2004ApJ...608..480B} around NML Cyg. They
assumed the star is near the peak intensity in the optical images and
coincident with the position of the strongest \hho\ maser feature, then
the maser shows an one-sided distribution within NML Cyg's nebula.  This
led to the suggestions that the masers are protected by the star's
envelope from Cyg OB2's radiation and the arc-like shape of
circumstellar material is the result of the interaction between the
molecular outflow from NML Cyg and the near-UV continuum flux from Cyg
OB2 \citep{2009ApJ...699.1423S}.  \citet{2008PASJ...60.1069N} presented
three-dimensional kinematic data for \hho\ maser features from
multi-epoch VLBI observations and suggested an expanding outflow with
the central star at the dynamical center.

However, after registering the radio photosphere to the SiO maser and
then to the \hho\ maser, we find that both the SiO maser and \hho\ maser
are highly asymmetrically distributed around the NML Cyg's central star,
which is located at southwest side of both the \hho\ and SiO maser
distribution, and there are \hho\ and SiO masers located north-west
of the central star. This is inconsistent with the model suggested by
\citet{2006AJ....131..603S,2009ApJ...699.1423S} based on the one-side
\hho\ maser distribution, since there are no masers located north-west
of the central star. And the central star is not near the center of the
two prominent maser clusters at \VLSR\ of about $-22$ and 5 \kms\ as
\citet{2008PASJ...60.1069N} suggested, but is closer to the maser
cluster at \VLSR\ of about $-22$ \kms, which is coincident with that
suggested by \citet{1996MNRAS.282..665R}.  Considering that kinematic
models for maser features are highly dependent on maser positions and
motions respective to the central star, our direct determination of
positions relative to the central star presents a crucial constraint for
understanding the properties of the circumstellar material surrounding
NML Cyg.

\section{Conclusions}

We have measured the trigonometric parallax and proper motion of NML Cyg
from multiple epoch VLBA observations of the circumstellar \hho\ and SiO
masers.  Both the distance and proper motion of NML Cyg are consistent
with those of the Cyg OB2 association within their joint uncertainties,
confirming that NML Cyg is associated with Cyg OB2. We revised the
stellar luminosity using the accurate distance from a trigonometric
parallax measurement, suggesting that the location of NML Cyg on the HR
diagram is consistent with the evolutionary track of an evolved star
with an initial mass of 25 \msun. However, the derived age of 8 Myr for
NML Cyg from the evolutionary track is much larger than that 2--3 Myr of
Cyg OB2, suggesting that the evolutionary track might not be
appropriate to NML Cyg.

After registering the radio photosphere to the \hho\ and SiO maser, we
find the central star is located close to a prominent \hho\ maser
cluster towards the north-west edge of the maser distribution.  This is
inconsistent with the model that the masers are off to one-side of the
star and also the model for an expanding outflow with the star in the
middle of two prominent \hho\ maser clusters as previously suggested.

\begin{acknowledgements}
We would like to thank the referee for very detailed and helpful
comments.  B. Zhang was supported by the National Science Foundation of
China under grant 11073046, 11133008.  A. Brunthaler was supported by a
Marie Curie Outgoing International Fellowship (FP7) of the European
Union (project number 275596).

\end{acknowledgements}

\clearpage


\clearpage

\onecolumn
\longtab{2}{
    \begin{longtable}{lcrrrrrr}
    \caption{Parallax and proper motion fits\label{tab:para_pm}} \\
\hline\hline
\tabincell{c}{Source \\ } &
\tabincell{c}{Region \\ } &
\tabincell{c}{\VLSR \\ (\kms) } &
\tabincell{c}{Parallax \\ (mas) } &
\tabincell{c}{\mux \\ (\masy) } &
\tabincell{c}{\muy \\ (\masy) } &
\tabincell{c}{\dx \\ (mas) } &
\tabincell{c}{\dy \\ (mas) } \\
\hline
\endfirsthead
\caption{Continuted.} \\
\hline\hline
\tabincell{c}{Source \\ } &
\tabincell{c}{Region \\ } &
\tabincell{c}{\VLSR \\ (\kms) } &
\tabincell{c}{Parallax \\ (mas) } &
\tabincell{c}{\mux \\ (\masy) } &
\tabincell{c}{\muy \\ (\masy) } &
\tabincell{c}{\dx \\ (mas) } &
\tabincell{c}{\dy \\ (mas) } \\
\hline
\endhead
\hline
\endfoot
\Jfour &B &   6.48   &  0.68 $\pm$ 0.10   &  $-$1.66 $\pm$ 0.58   &  $-$4.53 $\pm$ 0.19   &   $-$0.44 $\pm$ 0.19   &    2.19 $\pm$ 0.06   \\
    &   B &   6.06   &  0.72 $\pm$ 0.10   &  $-$1.70 $\pm$ 0.52   &  $-$4.33 $\pm$ 0.19   &   $-$0.37 $\pm$ 0.17   &    2.09 $\pm$ 0.05   \\
    &   B &   5.64   &  0.70 $\pm$ 0.09   &  $-$1.59 $\pm$ 0.48   &  $-$4.47 $\pm$ 0.18   &   $-$0.50 $\pm$ 0.16   &    2.18 $\pm$ 0.05   \\
    &   B &   5.22   &  0.70 $\pm$ 0.08   &  $-$1.68 $\pm$ 0.44   &  $-$4.55 $\pm$ 0.15   &   $-$0.63 $\pm$ 0.15   &    2.29 $\pm$ 0.04   \\
    &   B &   4.80   &  0.69 $\pm$ 0.06   &  $-$1.31 $\pm$ 0.47   &  $-$4.73 $\pm$ 0.12   &   $-$0.90 $\pm$ 0.16   &    2.55 $\pm$ 0.03   \\
    &   B &   4.37   &  0.55 $\pm$ 0.06   &  $-$1.85 $\pm$ 0.18   &  $-$5.50 $\pm$ 0.15   &   $-$0.66 $\pm$ 0.06   &    3.10 $\pm$ 0.04   \\
    &   B &   3.95   &  0.73 $\pm$ 0.07   &  $-$1.08 $\pm$ 0.52   &  $-$4.59 $\pm$ 0.14   &   $-$0.25 $\pm$ 0.17   &    3.80 $\pm$ 0.04   \\
    &   F & $-$18.80 &  0.69 $\pm$ 0.10   &  $-$1.81 $\pm$ 0.43   &  $-$5.25 $\pm$ 0.20   &  $-$85.02 $\pm$ 0.14   &    0.71 $\pm$ 0.06   \\
    &   F & $-$19.22 &  0.65 $\pm$ 0.12   &  $-$1.90 $\pm$ 0.39   &  $-$5.43 $\pm$ 0.25   &  $-$84.93 $\pm$ 0.13   &    0.94 $\pm$ 0.08   \\
    &   F & $-$19.64 &  0.88 $\pm$ 0.21   &  $-$3.34 $\pm$ 0.93   &  $-$5.01 $\pm$ 0.41   &  $-$83.88 $\pm$ 0.31   &    6.87 $\pm$ 0.12   \\
    &   G & $-$20.06 &  0.59 $\pm$ 0.12   &  $-$1.28 $\pm$ 0.33   &  $-$4.37 $\pm$ 0.32   &  $-$83.05 $\pm$ 0.11   &    6.98 $\pm$ 0.10   \\
    &   A & $-$20.48 &  0.66 $\pm$ 0.09   &  $-$1.99 $\pm$ 0.36   &  $-$4.85 $\pm$ 0.18   &  $-$83.15 $\pm$ 0.12   &    6.99 $\pm$ 0.05   \\
    &   A & $-$20.90 &  0.73 $\pm$ 0.10   &  $-$2.25 $\pm$ 0.45   &  $-$4.98 $\pm$ 0.19   &  $-$83.20 $\pm$ 0.15   &    6.96 $\pm$ 0.05   \\
    &   A & $-$21.32 &  0.66 $\pm$ 0.09   &  $-$1.24 $\pm$ 0.33   &  $-$5.06 $\pm$ 0.20   &  $-$76.67 $\pm$ 0.11   &    5.59 $\pm$ 0.06   \\
    &   A & $-$21.75 &  0.66 $\pm$ 0.10   &  $-$1.31 $\pm$ 0.32   &  $-$4.99 $\pm$ 0.22   &  $-$76.78 $\pm$ 0.11   &    5.63 $\pm$ 0.07   \\
    &   A & $-$22.17 &  0.75 $\pm$ 0.07   &  $-$2.23 $\pm$ 0.62   &  $-$5.15 $\pm$ 0.14   &  $-$84.58 $\pm$ 0.20   &    4.52 $\pm$ 0.04   \\
    &   A & $-$22.59 &  0.75 $\pm$ 0.08   &  $-$2.06 $\pm$ 0.54   &  $-$5.15 $\pm$ 0.14   &  $-$84.71 $\pm$ 0.18   &    4.50 $\pm$ 0.04   \\
    &   A & $-$23.01 &  0.70 $\pm$ 0.09   &  $-$2.12 $\pm$ 0.42   &  $-$5.17 $\pm$ 0.17   &  $-$84.77 $\pm$ 0.14   &    4.46 $\pm$ 0.05   \\
    &   R & $-$24.27 &  0.71 $\pm$ 0.06   &  $-$0.63 $\pm$ 0.53   &  $-$3.88 $\pm$ 0.12   &  $-$41.28 $\pm$ 0.18   &   33.57 $\pm$ 0.03   \\
    & & & & & & &\\
\Jsix & B &   6.48   &  0.63 $\pm$ 0.05   &  $-$1.56 $\pm$ 0.13   &  $-$4.29 $\pm$ 0.13   &   $-$0.23 $\pm$ 0.05   &    2.73 $\pm$ 0.04   \\
    &   B &   6.06   &  0.62 $\pm$ 0.01   &  $-$1.59 $\pm$ 0.02   &  $-$4.12 $\pm$ 0.16   &   $-$0.16 $\pm$ 0.01   &    2.64 $\pm$ 0.05   \\
    &   B &   5.64   &  0.64 $\pm$ 0.01   &  $-$1.50 $\pm$ 0.02   &  $-$4.24 $\pm$ 0.15   &   $-$0.29 $\pm$ 0.01   &    2.72 $\pm$ 0.05   \\
    &   B &   5.22   &  0.69 $\pm$ 0.01   &  $-$1.60 $\pm$ 0.02   &  $-$4.27 $\pm$ 0.17   &   $-$0.42 $\pm$ 0.01   &    2.83 $\pm$ 0.06   \\
    &   B &   4.80   &  0.61 $\pm$ 0.01   &  $-$1.21 $\pm$ 0.02   &  $-$4.53 $\pm$ 0.22   &   $-$0.68 $\pm$ 0.01   &    3.10 $\pm$ 0.07   \\
    &   B &   4.37   &  0.57 $\pm$ 0.09   &  $-$1.77 $\pm$ 0.26   &  $-$5.20 $\pm$ 0.22   &   $-$0.46 $\pm$ 0.09   &    3.64 $\pm$ 0.07   \\
    &   B &   3.95   &  0.56 $\pm$ 0.03   &  $-$0.96 $\pm$ 0.06   &  $-$4.46 $\pm$ 0.25   &   $-$0.02 $\pm$ 0.02   &    4.36 $\pm$ 0.08   \\
    &   F & $-$18.80 &  0.62 $\pm$ 0.03   &  $-$1.72 $\pm$ 0.08   &  $-$5.03 $\pm$ 0.18   &  $-$84.81 $\pm$ 0.03   &    1.26 $\pm$ 0.06   \\
    &   F & $-$19.22 &  0.64 $\pm$ 0.03   &  $-$1.82 $\pm$ 0.07   &  $-$5.14 $\pm$ 0.07   &  $-$84.73 $\pm$ 0.02   &    1.49 $\pm$ 0.02   \\
    &   F & $-$19.64 &  0.81 $\pm$ 0.20   &  $-$3.24 $\pm$ 0.50   &  $-$4.77 $\pm$ 0.59   &  $-$83.67 $\pm$ 0.17   &    7.42 $\pm$ 0.19   \\
    &   G & $-$20.06 &  0.61 $\pm$ 0.10   &  $-$1.21 $\pm$ 0.28   &  $-$4.04 $\pm$ 0.26   &  $-$82.85 $\pm$ 0.09   &    7.52 $\pm$ 0.08   \\
    &   A & $-$20.48 &  0.65 $\pm$ 0.04   &  $-$1.91 $\pm$ 0.11   &  $-$4.57 $\pm$ 0.15   &  $-$82.94 $\pm$ 0.04   &    7.54 $\pm$ 0.05   \\
    &   A & $-$20.90 &  0.67 $\pm$ 0.02   &  $-$2.15 $\pm$ 0.04   &  $-$4.75 $\pm$ 0.17   &  $-$82.99 $\pm$ 0.02   &    7.51 $\pm$ 0.06   \\
    &   A & $-$21.32 &  0.64 $\pm$ 0.04   &  $-$1.16 $\pm$ 0.10   &  $-$4.78 $\pm$ 0.21   &  $-$76.47 $\pm$ 0.03   &    6.14 $\pm$ 0.07   \\
    &   A & $-$21.75 &  0.64 $\pm$ 0.05   &  $-$1.23 $\pm$ 0.11   &  $-$4.71 $\pm$ 0.19   &  $-$76.57 $\pm$ 0.04   &    6.17 $\pm$ 0.06   \\
    &   A & $-$22.17 &  0.53 $\pm$ 0.01   &  $-$2.10 $\pm$ 0.02   &  $-$5.07 $\pm$ 0.20   &  $-$84.34 $\pm$ 0.01   &    5.08 $\pm$ 0.06   \\
    &   A & $-$22.59 &  0.58 $\pm$ 0.01   &  $-$1.95 $\pm$ 0.02   &  $-$5.02 $\pm$ 0.18   &  $-$84.48 $\pm$ 0.01   &    5.05 $\pm$ 0.06   \\
    &   A & $-$23.01 &  0.63 $\pm$ 0.09   &  $-$2.03 $\pm$ 0.38   &  $-$4.94 $\pm$ 0.17   &  $-$84.56 $\pm$ 0.13   &    5.01 $\pm$ 0.05   \\
    &   R & $-$24.27 &  0.54 $\pm$ 0.02   &  $-$0.52 $\pm$ 0.05   &  $-$3.75 $\pm$ 0.28   &  $-$41.05 $\pm$ 0.02   &   34.13 $\pm$ 0.09   \\
    & & & & & & &\\
\Jnine &B &   6.48   &  0.62 $\pm$ 0.02   &  $-$1.48 $\pm$ 0.07   &  $-$4.03 $\pm$ 0.05   &    0.13 $\pm$ 0.02     &    2.15 $\pm$ 0.02   \\
    &   B &   6.06   &  0.63 $\pm$ 0.03   &  $-$1.52 $\pm$ 0.08   &  $-$3.82 $\pm$ 0.13   &    0.20 $\pm$ 0.03     &    2.05 $\pm$ 0.04   \\
    &   B &   5.64   &  0.64 $\pm$ 0.04   &  $-$1.42 $\pm$ 0.10   &  $-$3.95 $\pm$ 0.12   &    0.07 $\pm$ 0.03     &    2.13 $\pm$ 0.04   \\
    &   B &   5.22   &  0.68 $\pm$ 0.05   &  $-$1.52 $\pm$ 0.11   &  $-$3.99 $\pm$ 0.16   &   $-$0.06 $\pm$ 0.04   &    2.24 $\pm$ 0.05   \\
    &   B &   4.80   &  0.62 $\pm$ 0.04   &  $-$1.14 $\pm$ 0.09   &  $-$4.23 $\pm$ 0.18   &   $-$0.33 $\pm$ 0.03   &    2.51 $\pm$ 0.06   \\
    &   B &   4.37   &  0.52 $\pm$ 0.07   &  $-$1.68 $\pm$ 0.35   &  $-$4.95 $\pm$ 0.14   &   $-$0.09 $\pm$ 0.11   &    3.06 $\pm$ 0.04   \\
    &   B &   3.95   &  0.59 $\pm$ 0.06   &  $-$0.89 $\pm$ 0.15   &  $-$4.14 $\pm$ 0.19   &    0.33 $\pm$ 0.05     &    3.77 $\pm$ 0.06   \\
    &   F & $-$18.80 &  0.64 $\pm$ 0.06   &  $-$1.64 $\pm$ 0.17   &  $-$4.72 $\pm$ 0.13   &  $-$84.45 $\pm$ 0.06   &    0.67 $\pm$ 0.04   \\
    &   F & $-$19.22 &  0.62 $\pm$ 0.01   &  $-$1.73 $\pm$ 0.13   &  $-$4.86 $\pm$ 0.03   &  $-$84.37 $\pm$ 0.04   &    0.90 $\pm$ 0.01   \\
    &   F & $-$19.64 &  0.81 $\pm$ 0.17   &  $-$3.17 $\pm$ 0.42   &  $-$4.47 $\pm$ 0.53   &  $-$83.31 $\pm$ 0.14   &    6.83 $\pm$ 0.17   \\
    &   G & $-$20.06 &  0.62 $\pm$ 0.10   &  $-$1.13 $\pm$ 0.27   &  $-$3.74 $\pm$ 0.26   &  $-$82.49 $\pm$ 0.09   &    6.93 $\pm$ 0.08   \\
    &   A & $-$20.48 &  0.62 $\pm$ 0.04   &  $-$1.83 $\pm$ 0.16   &  $-$4.32 $\pm$ 0.08   &  $-$82.58 $\pm$ 0.05   &    6.95 $\pm$ 0.02   \\
    &   A & $-$20.90 &  0.67 $\pm$ 0.04   &  $-$2.08 $\pm$ 0.10   &  $-$4.45 $\pm$ 0.12   &  $-$82.63 $\pm$ 0.03   &    6.92 $\pm$ 0.04   \\
    &   A & $-$21.32 &  0.64 $\pm$ 0.06   &  $-$1.08 $\pm$ 0.15   &  $-$4.49 $\pm$ 0.15   &  $-$76.11 $\pm$ 0.05   &    5.55 $\pm$ 0.05   \\
    &   A & $-$21.75 &  0.65 $\pm$ 0.06   &  $-$1.15 $\pm$ 0.17   &  $-$4.41 $\pm$ 0.14   &  $-$76.21 $\pm$ 0.06   &    5.58 $\pm$ 0.04   \\
    &   A & $-$22.17 &  0.55 $\pm$ 0.05   &  $-$2.03 $\pm$ 0.11   &  $-$4.76 $\pm$ 0.21   &  $-$83.99 $\pm$ 0.04   &    4.49 $\pm$ 0.07   \\
    &   A & $-$22.59 &  0.60 $\pm$ 0.05   &  $-$1.87 $\pm$ 0.10   &  $-$4.70 $\pm$ 0.18   &  $-$84.13 $\pm$ 0.04   &    4.46 $\pm$ 0.06   \\
    &   A & $-$23.01 &  0.65 $\pm$ 0.05   &  $-$1.95 $\pm$ 0.47   &  $-$4.64 $\pm$ 0.09   &  $-$84.20 $\pm$ 0.15   &    4.42 $\pm$ 0.03   \\
    &   R & $-$24.27 &  0.56 $\pm$ 0.06   &  $-$0.44 $\pm$ 0.13   &  $-$3.44 $\pm$ 0.22   &  $-$40.70 $\pm$ 0.05   &   33.54 $\pm$ 0.07   \\
    & & & & & & & \\
Combined & & &  & & & & \\
    &   B &   6.48   &                    &  $-$1.56 $\pm$  0.13  &  $-$4.29 $\pm$  0.11  &  &  \\
    &   B &   6.06   &                    &  $-$1.59 $\pm$  0.13  &  $-$4.11 $\pm$  0.11  &  &  \\
    &   B &   5.64   &                    &  $-$1.49 $\pm$  0.13  &  $-$4.24 $\pm$  0.11  &  &  \\
    &   B &   5.22   &                    &  $-$1.58 $\pm$  0.13  &  $-$4.33 $\pm$  0.11  &  &  \\
    &   B &   4.80   &                    &  $-$1.21 $\pm$  0.13  &  $-$4.50 $\pm$  0.11  &  &  \\
    &   B &   4.37   &                    &  $-$1.78 $\pm$  0.13  &  $-$5.14 $\pm$  0.11  &  &  \\
    &   B &   3.95   &                    &  $-$0.97 $\pm$  0.13  &  $-$4.39 $\pm$  0.11  &  &  \\
    &   F & $-$18.80 &                    &  $-$1.71 $\pm$  0.13  &  $-$5.02 $\pm$  0.11  &  &  \\
    &   F & $-$19.22 &                    &  $-$1.81 $\pm$  0.13  &  $-$5.14 $\pm$  0.11  &  &  \\
    &   G & $-$20.06 &                    &  $-$1.21 $\pm$  0.13  &  $-$4.02 $\pm$  0.11  &  &  \\
    &   A & $-$20.48 &                    &  $-$1.90 $\pm$  0.13  &  $-$4.59 $\pm$  0.11  &  &  \\
    &   A & $-$20.90 &                    &  $-$2.14 $\pm$  0.13  &  $-$4.78 $\pm$  0.11  &  &  \\
    &   A & $-$21.32 &                    &  $-$1.15 $\pm$  0.13  &  $-$4.79 $\pm$  0.11  &  &  \\
    &   A & $-$21.75 &                    &  $-$1.22 $\pm$  0.13  &  $-$4.71 $\pm$  0.11  &  &  \\
    &   A & $-$22.17 &                    &  $-$2.11 $\pm$  0.13  &  $-$4.98 $\pm$  0.11  &  &  \\
    &   A & $-$22.59 &                    &  $-$1.95 $\pm$  0.13  &  $-$4.97 $\pm$  0.11  &  &  \\
    &   A & $-$23.01 &                    &  $-$2.03 $\pm$  0.13  &  $-$4.94 $\pm$  0.11  &  &  \\
    &   R & $-$24.27 &                    &  $-$0.53 $\pm$  0.13  &  $-$3.66 $\pm$  0.11  &  &  \\
    &     &         & 0.620 $\pm$ 0.047   & & & & \\
\end{longtable}
\tablefoot{
Absolute proper motions are defined as $\mux = \mu_{\alpha \cos{\delta}}$ and $\muy = \mu_{\delta}$.
}
} 

\end{document}